%% file: Insights_into_the_Periodic_Gust_Response_of_Airfoils.tex
\documentclass{jfm}
\usepackage{graphicx}
\usepackage{epstopdf, epsfig}
\usepackage[]{subcaption} 
\usepackage{color}

\newcommand{\alphaG}{\hat{\alpha}_g}
\newcommand{\thetaH}{\hat{\theta}}

\shorttitle{Insights into the Periodic Gust Response of Airfoils}
\shortauthor{N. J. Wei et al.}

\title{Insights into the Periodic Gust Response of Airfoils}

\author{Nathaniel J. Wei\aff{1,3},
  Johannes Kissing\aff{1},
  Tom T. B. Wester\aff{2},
  Sebastian Wegt\aff{1},
  Klaus Schiffmann\aff{1},
  Suad Jakirlic\aff{1},
  Michael H{\"o}lling\aff{2},
  Joachim Peinke\aff{2},
   \and Cameron Tropea\aff{1}
   \corresp{\email{tropea@sla.tu-darmstadt.de}}}

\affiliation{\aff{1}Institute for Fluid Mechanics and Aerodynamics, Technische Universit\"at Darmstadt, Flughafenstra\ss e 19, 64347 Griesheim, Germany 
\aff{2}ForWind, Institute of Physics, University of Oldenburg, K\"upkersweg 70, 26129 Oldenburg, Germany
\aff{3}Department of Mechanical Engineering, Stanford University, Stanford, California 94305, USA}

\begin{document}

\maketitle

\begin{abstract}

The unsteady lift response of an airfoil in a sinusoidal gust can be modeled by two transfer functions: the first-order Sears function and the second-order Atassi function, albeit leading to different results under certain conditions. Previous studies have shown that the Sears function holds in experiments, but recently \cite{cordes_note_2017} reported experimental data that corresponded to the Atassi function rather than the Sears function. In order to clarify the observed discrepancy, the specific differences between these models are isolated analytically in this study and are related to physical gust parameters. Gusts with these parameters are then produced in wind-tunnel experiments using an active-grid gust generator. Measurements of the unsteady gust loads on an airfoil in the wind tunnel at Reynolds numbers ($\rm{Re_c}$) of $2.0 \times 10^5$ and $2.6 \times 10^5$ and reduced frequencies between $0.09$ and $0.42$ confirm that the decisive difference between the Sears and Atassi functions lies in the character of the gust and not in the characteristics of the airfoil. The differences in the gust-response data between Sears and Atassi gust conditions are shown to be significant only at low reduced frequencies. These findings are supported by numerical simulations of the experimental setup. Finally, the influence of boundary-layer turbulence on experimental convergence with model predictions is investigated. These results serve to clarify the conditions under which the Sears and Atassi functions can be applied, and they establish the validity of both in an experimental context.

\end{abstract}

\begin{keywords}

\end{keywords}

\input{01_introduction}

\input{02_theoretical_considerations.tex}

\input{03_experimental_details.tex}

\input{04_experimental_considerations.tex}

\input{05_measurement_results.tex}

\input{06_discussion_and_conclusions.tex}

\input{07_acknowledgements.tex}

\bibliographystyle{jfm}
\bibliography{SearsAtassi}

\end{document}

%% file: 01_introduction.tex
\section{Introduction}

Throughout the history of modern aerodynamics, wind gusts have consistently proved to be a challenge for theoreticians and engineers alike. The unsteady loads on an airfoil encountering a periodic gust can be critical to the safety and performance of piloted fixed-wing aircraft. These gust effects are similarly problematic at smaller scales, such as for flying animals and micro-air vehicles \cite[e.g.][]{reynolds_wing_2014, zarovy_experimental_2010}. In addition, atmospheric gusts can greatly reduce the effective life span of the rotating blades on rotorcraft and wind turbines \cite[][]{spinato2009reliability}. Even bridges have been known to fail under heavy gust-loading conditions \cite[e.g.][]{billah_resonance_1991}. Thus, for engineers attempting to account for gust-loading effects when designing aerodynamic components, an accurate understanding of the effect of gusts on structures is essential.

In order to better understand and capture the dynamics of gust-airfoil interactions, \cite{sears_systematic_1938} modeled the problem using the principles of thin-airfoil theory. By approximating the airfoil as a series of point vortices distributed along its camber line, potential-flow analysis can be applied. When a periodic gust normal to the airfoil is supplied as an inflow condition, the effects on the vorticity profile on and in the wake of the airfoil can be computed. Integrating the vorticity profile on the airfoil yields what Sears called the lift contribution of the apparent mass \cite[cf.][]{von_karman_airfoil_1938, sears_aspects_1941}, whereas integrating the vorticity profile in the wake provides the forces due to unsteady effects in the wake itself. The latter component of the lift force involves a function $C(k)$ developed by \cite{theodorsen_general_1934}, which depends only on the reduced frequency $k$ of the airfoil. The reduced frequency is defined  by \cite{leishman_principles_2006} as 

\begin{equation}
    k = \pi \frac{f c}{U_\infty},
\end{equation}

\noindent where $f$ is the frequency of the unsteady inflow in Hertz, $c$ is the chord length of the airfoil, and $U_\infty$ represents the free-stream velocity. A transfer function for the unsteady lift on an airfoil in a periodic gust can thus be constructed by combining the apparent-mass and wake-effect lift components with the quasi-steady lift derived from the Kutta-Joukowski theorem. This transfer function is known as the Sears function. While the function is restricted by its derivation to flows in which the assumptions of thin-airfoil theory and potential flow hold, it is nevertheless attractive because of its simplicity: like the aforementioned Theodorsen function, it only depends on the reduced frequency of the incoming gust, and is mathematically straightforward. In addition to being widely used to predict unsteady loads in gust-response problems, the Sears function has also enjoyed broader applicability across research areas ranging from active flow control \cite[e.g.][]{zhao_buffet_2016} to aeroacoustics \cite[e.g.][]{wang_large-eddy_2017}.

The Sears function was extended by \cite{goldstein_complete_1976} and \cite{atassi_sears_1984} using second-order models to account for the distortion of the gust and flow field due to the presence of the airfoil. This allowed the effects of airfoil camber and mean angle of attack to be incorporated. In the process, the possibility of accounting for fluctuations in the horizontal as well as vertical velocity was introduced. This second-order model will be referred to as the Atassi function. \cite{lysak_prediction_2013} developed a correction for the Sears function that takes effect at high reduced frequencies ($k > 1$), and \cite{massaro_effect_2015} derived an additional correction for three-dimensional effects as a function of aspect ratio. It is worth noting that these analytical extensions do not deviate from the Sears function itself under its original assumptions and flow conditions.

Despite its age, applicability, and adaptability, the Sears function has not been extensively validated in experiments, due in part to the difficulty of creating repeatable sinusoidal gusts in the vertical velocity component only. Early attempts, such as that of \cite{hakkinen_theoretical_1957}, proved inconclusive because of limitations in measurement technology. \cite{commerford_unsteady_1974} tested the function on a thin plate at only a single reduced frequency ($k = 3.9$), and found significant deviation from the Sears prediction. In contrast, \cite{jancauskas_aerodynamic_1986} used two controlled-circulation airfoils to produce single-frequency sinusoidal gusts over a range of reduced frequencies up to $k = 0.37$, and found excellent agreement with the Sears function when a NACA-0006 airfoil was used as the test profile. Until recently, this was the only study in the literature to use single-frequency sinusoidal gusts to test the Sears function experimentally. \cite{sankaran_direct_1992} showed that the Sears function could also be reproduced using a spectral decomposition of the force signal to isolate various reduced frequencies within multimodal gusts, such as those occurring in the flow behind a turbulence-generating screen. This spectral approach was used by \cite{larose_experimental_1999} and \cite{hatanaka_new_2002}, who both observed significant deviation from the Sears function when cross-sectional profiles of various bridge decks were employed as the test profile. The latter study reproduced the general trend of the Sears function in experiments with a NACA-0012 airfoil. More recently, \cite{lysak_measurement_2016} were able to verify their thickness-correction model for a range of airfoils using the same kind of measurement technique.

The Sears function has its roots in potential-flow theory and therefore the use of turbulent wind tunnel flows to verify the theory creates ambiguities when comparing the inflow conditions to the original conditions of the theory. These ambiguities affect the applicability of the Sears function: \cite{jancauskas_aerodynamic_1983} showed that increasing the background turbulence of the incoming gust flow could improve experimental correspondence with the Sears function on bridge-deck profiles that deviated significantly from  theory under more laminar conditions. To clarify this, \cite{cordes_note_2017} returned to the generation of single-frequency sinusoidal gusts in order to test whether the Sears function would hold for a Clark-Y airfoil. In contrast with what one might expect from the results of \cite{lysak_measurement_2016}, a trend opposite to that of Sears was found, which matched instead the trend of the Atassi function. This difference was attributed to the distortion of the flow field due to airfoil angle of attack and camber that is modeled by Atassi but not by Sears. However, numerical simulations of the experimental apparatus by \cite{wegt_numerische_2017} showed that the active grid used to generate sinusoidal gusts for the experiments produced highly three-dimensional flows and that the relative strength of the gusts were not constant over the range of tested reduced frequencies. This has been subsequently confirmed in experiments by \cite{traphanaerodynamic}. Thus, while the results of \cite{cordes_note_2017} show that there are differences between the Sears and Atassi functions that delineate where each can be applied, they remain ambiguous as to whether these differences stem from the choice of airfoil or the inflow conditions themselves.

In order to determine more precisely the conditions under which these analytical transfer functions can be applied in real-world flow situations, it is necessary to disentangle the differences between the Sears and Atassi functions, both in their formulation and in their correspondence with experimental data. Therefore, in this paper we seek to clarify these distinctions by validating both functions under their respective flow conditions. We isolate differences within the theories themselves in terms of the scenarios they describe and the mathematical conventions they rely on, and we demonstrate how these differences influence the trends seen in experimental data. A more precise understanding of both theories and their performance in experiments  provide a foundation for interpreting previous experimental studies, and will establish more clearly the conditions under which these theories can be applied in practice.

%% file: 02_theoretical_considerations.tex
\section{Theoretical considerations}

In this section, the theories of Sears and Atassi are analyzed and compared in order to demonstrate their essential differences. While the Atassi function is derived using the same principles as the Sears function, small differences in its formulation lead to significant differences in its properties and behavior.

\subsection{Sears and Atassi formulations}

Both transfer functions relate the unsteady lift $L_{dyn}$ from the gust-airfoil interaction to the quasi-steady lift $L_{qs}$, or the lift amplitude that would result from the same range of gust angles of attack in steady flow. They take the form

\begin{equation}
    h_L = \frac{L_{dyn}}{L_{qs}} = \frac{\hat{L}_{dyn}}{\hat{L}_{qs}} e^{i \phi},
    \label{eqn:tf}
\end{equation}
where hats denote amplitudes and $\phi$ represents the phase shift between the dynamic and quasi-steady lift-force signals. The time phase of the gust is taken to be positive, in accordance with \cite{atassi_sears_1984}. The dynamic lift in the Sears function is computed as a combination of quasi-steady, apparent-mass, and wake-circulation lift components. The Atassi function adds contributions from distortions of the flow field due to camber and mean angle of attack to the dynamic lift from the Sears function. Thus,  the two theories are quite similar; the Atassi function is simply an extension of the Sears function.

\subsection{Influence of the streamwise reduced frequency $k_{2}$}

\begin{figure}
  \centerline{\includegraphics[width=0.8\textwidth]{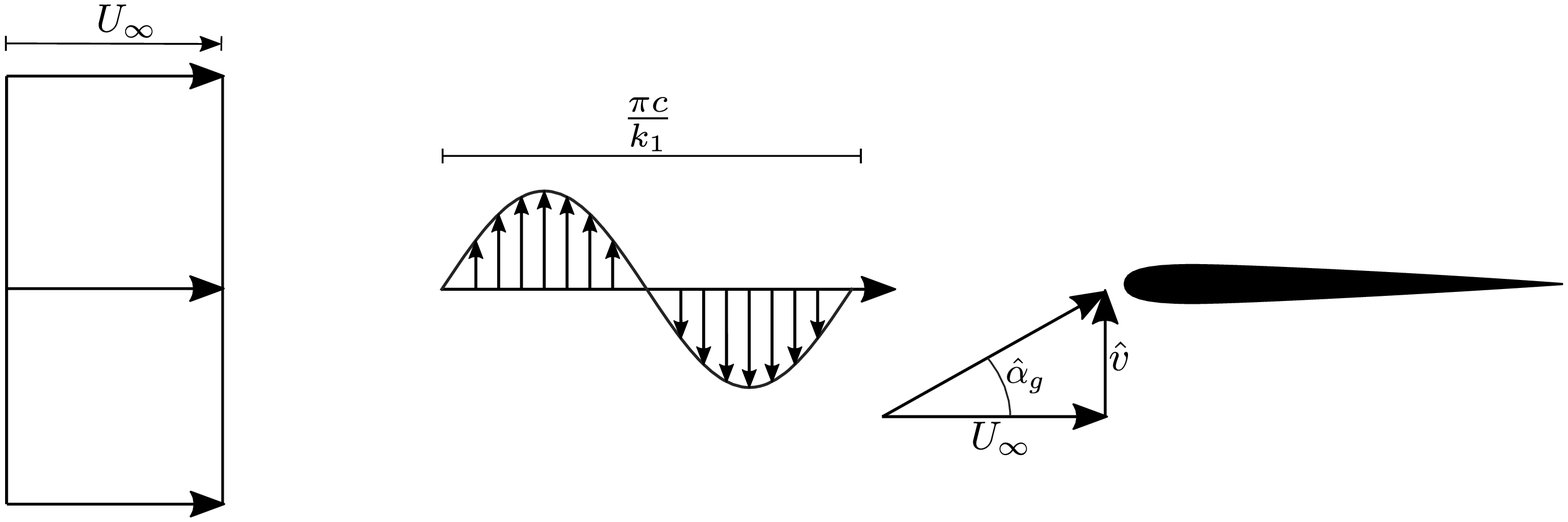}}
  \caption{Inflow conditions according to Sears. Fluctuations in the normal velocity component $\hat{v}$ with reduced frequency $k_1$ impinge on the airfoil. The resulting angle-of-attack variations are represented by $\alphaG$.}
\label{fig:InflowSears}
\end{figure}

\begin{figure}
  \centerline{\includegraphics[width=0.8\textwidth]{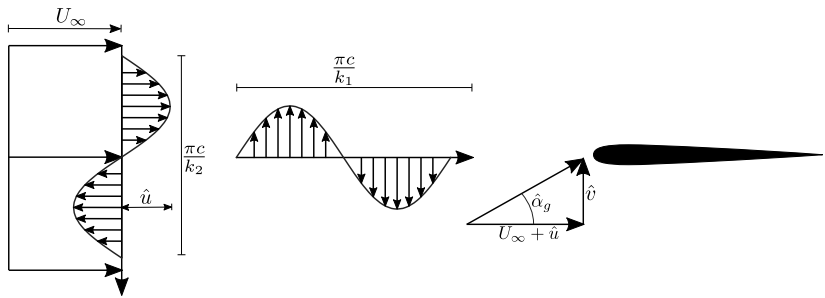}}
  \caption{Inflow conditions according to Atassi. In addition to the normal-velocity fluctuations from the Sears problem, streamwise-velocity fluctuations with amplitude $\hat{u}$ and reduced frequency $k_2$ are also present. This extra gust profile is mathematically coupled with the Sears-problem gust, thereby complicating the Atassi formulation of the problem.}
\label{fig:InflowAtassi}
\end{figure}

The most obvious difference between the theories of Sears and Atassi lies in the setup of the problem to be modeled. The Sears function involves only periodic velocity fluctuations in the component normal to the airfoil  ($v$), as shown in figure \ref{fig:InflowSears}. The theory of Atassi, in contrast, includes periodic velocity fluctuations in the streamwise velocity component $u$ as well. These fluctuations, shown in figure \ref{fig:InflowAtassi}, are characterized according to Atassi's notation by the reduced frequency $k_2$, and the vertical-velocity oscillations corresponding to those of the theory of Sears by the reduced frequency $k_1$. The waveform of these streamwise fluctuations runs normal to the airfoil at a velocity of $u = \pi\frac{fc}{k_2}$ (m/s), so that the streamwise velocity at any height in the flow fluctuates about the free-stream velocity with an amplitude of $\hat{u}$. When $k_2$ is zero, the streamwise velocity fluctuations are not present, and the Atassi problem reduces to that of Sears.


This is made explicit in the mathematical definition of the Atassi function. If the Sears function is denoted as $S(k_1)$, the Atassi function can be written as

\begin{equation}
    h_{L}(k_1, k_2) = \frac{k_1}{\left|k\right|} S(k_1) + \alpha_\beta h_\beta(k_1, k_2) + \alpha_m h_m(k_1, k_2),
    \label{eqn:Atassi}
\end{equation}

where $\left|k\right| = \sqrt{{k_1}^2+{k_2}^2}$, $\alpha_\beta$ is the mean angle of attack of the airfoil, and $\alpha_m$ is the airfoil camber. $h_\beta$ and $h_m$ are both relatively complicated functions of $k_1$ and $k_2$ and are defined in \cite{atassi_sears_1984}. Both functions become zero when $k_2$ is zero. From this definition, it is evident that the Atassi function is distinct from the Sears function if and only if $k_2 \neq 0$. For the purposes of this work, only airfoils at zero mean angle of attack ($\alpha_\beta = 0$) and with zero camber ($\alpha_m = 0$) are considered, so that these terms do not contribute to the Atassi function $h_L$.

\begin{figure}
  \centerline{\includegraphics[width=0.5\textwidth]{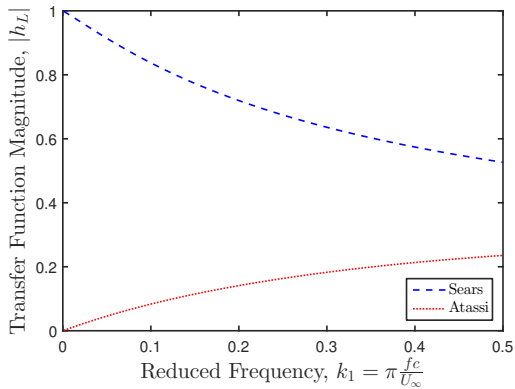}}
  \caption{Profiles of the Sears and Atassi functions. The Atassi curve is given for $k_2 = 1$. When $k_2$ is zero, the Atassi function collapses onto the Sears function. The phase remains the same irrespective of the value of $k_2$.}
\label{fig:SearsAtassiCurves}
\end{figure}

The influence of $k_2$ is clearly visible in the magnitudes of the resulting transfer functions, shown in figure \ref{fig:SearsAtassiCurves}. When $k_2$ takes on a constant nonzero value, the Atassi function displays an inverted trend with regard to the Sears function: rather than starting from unity and decreasing monotonically, the function starts from the origin and increases. At high reduced frequencies, both functions converge asymptotically onto the same trajectory. The addition of the camber correction term $h_m$ serves only to shift the curve along the reduced-frequency axis. Similarly, the angle-of-attack correction term $h_\beta$ causes the curve to rise more sharply at low reduced frequencies before falling asymptotically to match the Sears curve. For the purposes of this study,  the importance of these correction terms is that neither is responsible for producing the inverted trend of the Atassi curve in relation to the Sears curve. This emphasizes the point that the fundamental difference between the theories of Sears and Atassi lies not in the corrections for flow-field distortion in the presence of the airfoil, as \cite{cordes_note_2017} hypothesized, but rather in the parameter $k_2$.

\subsection{Transfer function normalization}

The importance of the $k_2$ parameter becomes clearer when the normalizations of the two transfer functions are scrutinized. These factors are not used to compute the transfer functions themselves, but rather to normalize the unsteady lift-force data obtained from gust-response experiments ($\hat{L}_{dyn}$). They are therefore measures of the quasi-steady lift ($\hat{L}_{qs}$), which by definition must include a parameter to represent the strength of the unsteady gust fluctuations. For the Sears function, the factor is given by the relation

\begin{equation}
    \hat{L}_{qs} = \pi \rho c U_\infty \hat{v} \approx \pi \rho c {U_\infty}^2 \hat{\alpha}_G,
    \label{eqn:NormSears}
\end{equation}

\noindent where $\hat{v}$ is the amplitude of the velocity fluctuations normal to the airfoil. Since $\hat{v} << U_\infty$, $\hat{v}$ can be approximated as being directly proportional to $\alphaG$, the amplitude of the inflow angles of attack created by the gust. From this normalization, it is clear that the Sears function should be independent of the gust strength $\hat{v}$ and thus the gust-angle amplitude $\alphaG$.

Defining a gust strength is made more difficult when the streamwise gust defined by $k_2$ is introduced. In Atassi's theory, a single parameter $\epsilon$ represents simultaneously the gust strength of the normal and streamwise velocity fluctuations. The amplitudes of the gusts in the streamwise and normal velocity components are given respectively as $\hat{u} = U_\infty \epsilon k_2 / \left|k\right|$ and $\hat{v} = U_\infty \epsilon k_1 / \left|k\right|$ (cf.\ figure \ref{fig:InflowAtassi}). The gust strength is defined such that $\epsilon << 1$, implying that $\hat{u} << U_\infty$ and $\hat{v} << U_\infty$. This allows the gust strengths from the theories of Sears and Atassi to be compared via the physical quantity $\alphaG$. The gust-angle amplitude can thus be represented as

\begin{equation}
    \hat{\alpha}_G = \arctan\left(\frac{\hat{v}}{U_\infty+\hat{u}}\right) \approx \arctan\left(\frac{\hat{v}}{U_\infty}\right) \approx \frac{\epsilon k_1}{\sqrt{{k_1}^2+{k_2}^2}}.
    \label{eqn:AtassiAlphaG}
\end{equation}

When $k_2$ is zero, the gust strength $\epsilon$ is simply $\alphaG$, as for the Sears function. When $k_2$ takes on a nonzero value, however, a constant value of $\epsilon$ means that $\alphaG$ becomes a monotonically increasing function of $k_1$. Thus, the gust strength in the Atassi problem is not strictly tied to the physical quantity $\alphaG$. This is important for the normalization of the Atassi function, given by

\begin{equation}
    \hat{L}_{qs} = \pi \rho c {U_\infty}^2 \epsilon.
    \label{eqn:NormAtassi}
\end{equation}

According to the relation given in equation \ref{eqn:AtassiAlphaG}, this term is a nontrivial function of $\alphaG$ and $k_2$ through the $\epsilon$ parameter's dependence on $k_1$. This is of no consequence to the shape of the transfer functions, as the normalization simply shows that the Atassi function, like the Sears function, remains independent of $\alphaG$. However, it does imply that if $\epsilon$ is to be held constant in experiments, $\alphaG$ cannot remain constant, but rather must be increased with $k_1$. More generally, if sensible values for the Atassi function are to be obtained from experiments, this analysis shows that either $\epsilon$ or $\alphaG$ and $k_2$ must be known for each investigated $k_1$.

The normalizations of the transfer functions thus allow a wide range of experimental conditions to be captured by a single curve defined only by $k_2$, provided that the parameters $k_2$, $\alphaG$, and by extension $\epsilon$ can be controlled and measured accurately.

\subsection{Significance for experiments}

The mathematical details outlined above are important considerations for the setup of experiments investigating the Sears and Atassi problems. They show that control over the fluctuations in the streamwise velocity is of paramount importance. Thus, to retrieve a Sears trend, $\hat{u}$ must be kept as small as possible so that $k_2$ can be set to zero. If this condition is not taken into account, then the resulting conditions will be those of the Atassi problem and not strictly those of Sears. No previous study has considered systematically controlling the oscillations in $u$ while performing Sears measurements, and therefore a central goal of this study is to demonstrate the importance of this distinction in an experimental context.

In addition, in order to compute  experimental transfer-function values from unsteady lift-force data, the parameters $\alphaG$ and $k_2$ must be accurately determined, and the correct normalizations for the transfer functions in question must be employed. In previous studies \cite[e.g.][]{cordes_note_2017}, the proper normalization for the transfer functions had not been clarified, and $\alphaG$ was not characterized over the entire range of experimental gust conditions. These are considerations that must be accounted for in experiments seeking to establish the difference between the two transfer functions.

The transfer-function phase profiles for the Sears and Atassi functions are identical when camber and mean angle of attack are both zero, irrespective of the values of $k_2$, $\epsilon$, and $\alphaG$. Therefore, the phase is not considered in detail here.

%% file: 03_experimental_details.tex
\section{Experimental details}

\subsection{Wind Tunnel and Grid}

\begin{figure}
  \centerline{\includegraphics[width=0.6\textwidth, trim=0 0 0 0,clip]{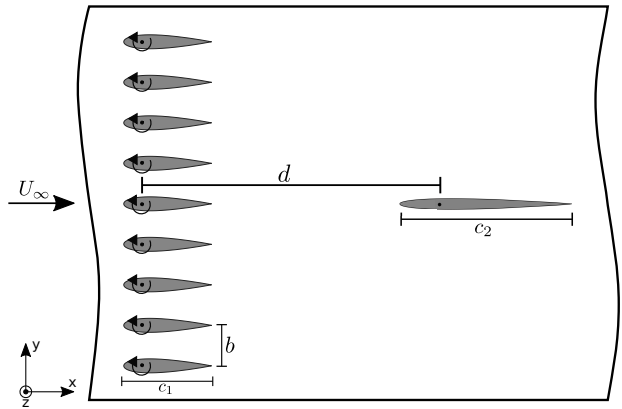}}
  \caption{Schematic showing the setup of the wind tunnel and active grid (illustration not to scale). The distance between the grid and the quarter-chord point of the test airfoil is $d = 1.1$ m. The vanes of the active grid have a chord length of $c_1 = 71$ mm.}
\label{fig:Setup}
\end{figure}

The experiments reported in this study were carried out in a closed-loop wind tunnel at the University of Oldenburg, details of which are given in \cite{knebel_atmospheric_2011}. The test section had a cross-sectional profile of $1.00 \times 0.80\;\mathrm{m^2}$. An active grid was installed in the tunnel at the front of the test section, $1.1\;\mathrm{m}$ upstream from the quarter-chord point of the airfoil. It was composed of nine 3D-printed vanes, each with a NACA 0016 cross-sectional profile with $71\;\mathrm{mm}$ chord and $0.8\;\mathrm{m}$ span. The turbulence intensity in the middle of the tunnel, computed from data from hot-wire anemometers, was less than $0.3\%$. The vanes were pitched about their quarter-chord point in sinusoidal profiles. The precise details of the motion protocols employed in this study are given below in section \ref{sec:grid_protocols}. A schematic of the wind tunnel and active grid is provided in figure \ref{fig:Setup}.

A NACA 0006 airfoil with a $202\;\mathrm{mm}$ chord and $800\;\mathrm{mm}$ span was constructed at the TU Darmstadt and served as a test airfoil. It was composed of two carbon-fiber shells attached to an aluminum spine, so that the airfoil was both smooth and rigid against torsion. This profile was selected to conform to the thin-airfoil assumption of Sears, based on predictions from the XFOIL software \cite[cf.][]{drela_xfoil:_1989} for the range of angles of attack the airfoil could encounter without incurring flow separation on the surface. A thin strip of fluorescent foil at the location of the laser sheet reduced surface reflections when particle image velocimetry (PIV) measurements were performed.

\subsection{Hot-wire measurements}
To obtain empirical relations between the grid amplitude $\hat{\theta}$ and gust-angle amplitude $\hat{\alpha}$ over the range of potential experimental flow conditions, as well as estimates for the Atassi gust parameters $\epsilon$ and $k_2$, a large parameter space of grid amplitudes and reduced frequencies was traversed for two distinct grid configurations (detailed in section \ref{sec:grid_protocols}). Data were collected using a CTA System from \textit{DANTEC DYNAMICS},  consisting of a \textit{StreamLine Automatic Calibrator} for X-wire calibration, a \textit{StreamLine} frame with two CTA modules, a $6 \rm{mm}$ short support of type \textit{55H24} and an X-wire of type \textit{55P51}. \\
During the characterization of the inflow the calibrated X-wire was placed in at the centerline of the empty wind tunnel at the intended leading edge position of the airfoil $1 \rm{m}$ downstream of the active grid. For each frequency and amplitude 40,000 values were taken at a sampling frequency of $20 \rm{kHz}$ resulting in a duration of $20 \rm{s}$ per grid protocol. \\
To cover a broad range of possible reduced frequencies and gust amplitudes, the grid was characterized for frequencies from $1 \rm{Hz}$ to $10 \rm{Hz}$ with grid amplitudes from $2.5 \rm{^\circ}$ up to $20 \rm{^\circ}$ for free stream velocities between 15 m/s and 25 m/s.

\subsection{Force Measurements}

A pair of K3D120 3-axis force balances and a TS110 moment sensor from \textit{ME-Messsysteme} allowed forces and moments to be measured at the airfoil's quarter-chord point. Thirty seconds of force data were collected for each measurement. All measurements, including measurements with the PIV system, were started simultaneously via a trigger provided two seconds after the active-grid protocol was started.

To compute transfer functions  from experimental data, the force data were first shifted from the airfoil frame of reference to that of the wind tunnel. The first two seconds of each data set were discarded in order to remove start-up effects. A Fourier transform of the signal yielded the magnitude of the lift force fluctuations, corresponding to $\hat{L}_{dyn}$ from equation \ref{eqn:tf}. For the Sears function, the quasi-steady lift amplitude $\hat{L}_{qs}$ was interpolated from steady lift curves $L_{U_\infty}$ using the relation

\begin{equation}
    \hat{L}_{qs} = f(U_\infty, \alphaG, \alpha_m) = \frac{1}{2}\big(L_{U_\infty}(\alpha_m + \alphaG) - L_{U_\infty}(\alpha_m - \alphaG)\big).
\end{equation}

Because $\alpha_m$ in these experiments was always zero, this was approximately equivalent to

\begin{equation}
    \hat{L}_{qs} = \frac{\partial L_{U_\infty}}{\partial \alpha} \alphaG,
\end{equation}

\noindent where $\frac{\partial L_{U_\infty}}{\partial \alpha}$ represents the lift slope of the airfoil as computed from a fit of the linear portion of the steady lift curve. This is the experimental equivalent of the ideal quasi-steady lift factor given in equation \ref{eqn:NormSears}, since a value for the lift slope that reflected the test airfoil used in experiments was preferable to the theoretical lift slope $\pi \rho c {U_\infty}^2$. This modified form was used to compute $\hat{L}_{qs}$ for the Atassi function, with $\epsilon$ substituted for $\alphaG$. $\epsilon$ was computed from equation \ref{eqn:AtassiAlphaG}.

The data points were validated using both force-balance data and, for the six cases in which stereo PIV was employed, phase-averaged velocity vector fields. Error bars were computed from the standard deviation of the phase-averaged lift-force signal and the standard deviation of the gust-angle amplitudes from hot-wire data. For data taken at $U_\infty = 10$ m/s, no hot-wire data was available, and therefore gust-angle amplitudes were extracted from the stereo PIV vector fields and errors were estimated.
	
It was expected that the Fourier transformed lift-force data would show a single large peak at the frequency of gust generation, but occasionally smaller peaks at other frequencies were observed. If any of these secondary peaks were over 20\% as large as the main peak, this suggested that the lift force was not purely sinusoidal. These data sets were then removed from the transfer function plot, so that the resulting data only represented single frequency sinusoidal gusts.

\subsection{PIV Setup}
To investigate the spatial/temporal structures occurring in the flow, additional PIV measurements were performed. The PIV setup was stereoscopic a setup from \textit{LaVision}, composed of two \textit{Phantom Miro 320S} high-speed cameras operated at a frame rate of $500 \rm{Hz}$. This reduced temporal resolution and a RAM size of the cameras of $12 \rm{GB}$ enabled a total measurement time of $5.5 \rm{s}$. The cameras were mounted above and below the test section. A \textit{Litron LDY-303HE} Nd:YLF dual-cavity high-speed pulsed laser illuminated the suction side of the airfoil from the rear of the tunnel. Images were recorded and evaluated with \textit{LaVision's DaVis 8.4} software with $16\times16$-pixel interrogation areas, 50\% overlap, and outlier interpolation.

\subsection{Numerical setup}
Numerical simulations of the flow field in the wind tunnel were performed with the aim of complementing the experimental measurements.  Two configurations were studied, corresponding to two different inflow conditions.  The first  configuration corresponded to conditions specified by Sears (see figure \ref{fig:InflowSears}) and simulated the flow past a NACA 0006 airfoil with chord length $c_2=0.2$ m. The second simulation involved only the nine oscillatory vanes without an airfoil model and was aimed at reproducing the tunnel unsteadiness as encountered in the experiments (see figure \ref{fig:Setup}).

The computations were performed within the Unsteady RANS (Reynolds-Averaged Navier-Stokes) computational framework employing the baseline and eddy-resolving versions of the near-wall Reynolds-stress-model, RSM, by \cite{Jakirlic2002}. The model is capable of asymptotically capturing both the strengthened viscosity effects and the kinematic wall blocking. The latter model property is expressed in terms of both Reynolds-stress and stress-dissipation anisotropy coping with correct asymptotic behavior of all turbulence quantities when approaching the  wall.  Unlike the baseline model, its eddy-resolving formulation (denoted  by Improved Instability-Sensitive RSM - IISRSM) is capable of resolving the fluctuating turbulence. This model feature, created in line with the Scale-Adaptive Simulation (SAS) methodology proposed originally by \cite{Menter2010} in conjunction with their $k-\omega$ SST eddy-viscosity model of turbulence, is achieved by suppressing  the modelled turbulence towards a corresponding sub-scale level. The characteristic size of the unresolved sub-scale turbulent eddies is proportional to the von Karman length scale (comprising the second derivative of the velocity field: $\nabla^2 \bf U$), which mimics, analogously to the grid-spacing $\Delta$ in LES (Large-Eddy Simulation), the relevant length-scale of the SAS-related residual turbulence. The  turbulence suppression in the SAS modelling framework is achieved by introducing an additional production term ($P_{SAS}$) into the scale-supplying equation governing the $\omega$-quantity (with $\omega \propto \varepsilon/k$ representing inverse turbulent time scale). Herewith, an appropriate enhancement of the turbulence activity originating from the resolved motion, especially within the shear layer regions, is ensured; interested readers are referred to \cite{Jakirlic2015} and \cite{Maduta2017} for more details. 

The model equations are implemented into the open source toolbox OpenFOAM\textsuperscript{\textregistered} with which all present simulations were performed. The code employs a finite-volume discretisation method based on the integral form of the general conservation law applied in conjunction with block-structured grid arrangements. The flow domains considered are meshed by the OpenFOAM\textsuperscript{\textregistered} relevant utility denoted as ``blockMesh''. The temporal resolution adopted guarantees a Courant number smaller than one in the entire solution domain. The discretisation of the convection terms is achieved using a blended central differencing scheme implemented in the differed-correction approach manner. This methodology is considered adequate in view of a fine grid used in the regions with high gradients. 

\begin{figure}
\centerline{\includegraphics[width=0.8\textwidth,clip]{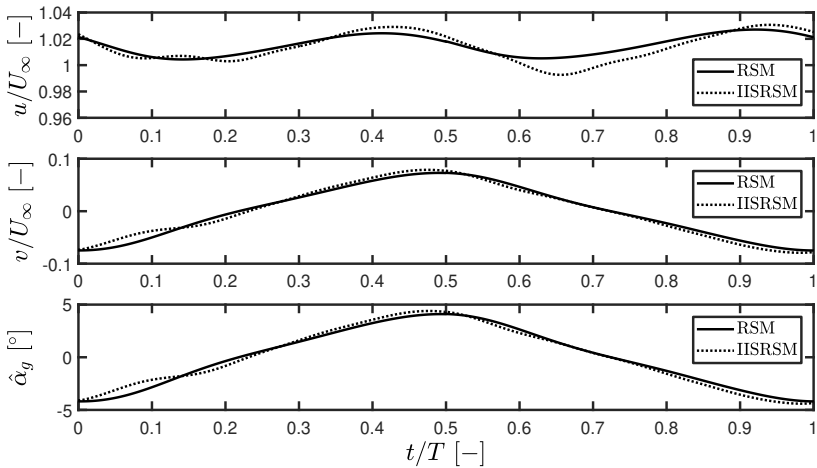}}
\caption{Computationally obtained time variation of the streamwise and normal component of the mean velocity field and gust angle amplitude by using baseline and eddy-resolving Reynolds stress models.}
\label{fig:RSM-IISRSM-Comparison}
\end{figure}

The first computational setup is concerned with the flow past a NACA 0006 airfoil. The inflow cross-section is positioned immediately downstream of the active grid. The incoming gusty flow conditions are represented by a time-dependent sinusoidal velocity condition at the inlet $($compare with figure \ref{fig:InflowSears}$)$. The two-dimensional solution domain (not shown here) with the inflow plane located at $1.1$ m upstream of the test airfoil’s quarter-chord point (figure~\ref{fig:Setup}) and the outlet plane located at $1.5$ m downstream of the airfoil’s quarter-chord point, is meshed with 59,300 hexahedral grid cells. The grid is squeezed with an adequate grading towards the airfoil surface. The governing equations are integrated to the walls; the wall-adjacent computational node at the airfoil is situated deep within the viscous sublayer, ensuring the value of the dimensionless wall distance $y^+ \leq 1$. Such a fine grid resolution is necessary for applying the exact wall boundary conditions, formulated in relation to the physically correct behavior of the mean flow and turbulent quantities.  The upper and lower bounding of the solution domain is represented by symmetry planes positioned equidistantly at $0.5$ m from the airfoil's chord line. Application of the symmetry conditions is justified as they do not influence the flow around the airfoil. These computations were performed employing the baseline RSM. The credibility of performing  two-dimensional computations is based on the experience gained by computing the flow in the empty plenum (the second configuration). These results,  obtained for both two-dimensional and three-dimensional computations, exhibit no noticeable difference (see figure \ref{fig:RSM-IISRSM-Comparison} and associated discussion). 

\begin{figure}
  \centerline{\includegraphics[width=0.8\textwidth,clip]{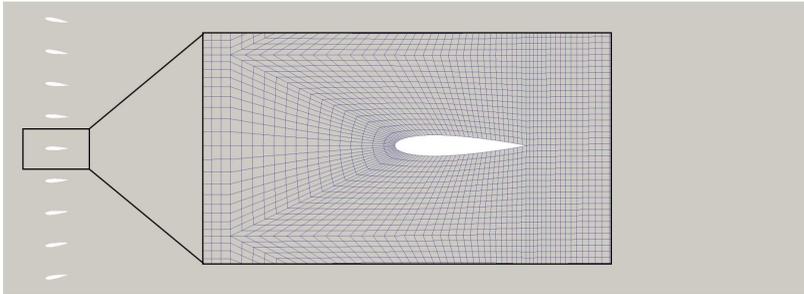}}
  \caption{Computational domain accommodating the active grid and a blow-up of the numerical grid meshing the region immediately surrounding a vane.}
\label{fig:Numeric_ActiveGrid}
\end{figure}

The second  configuration deals with the entire plenum (accounting also for the plenum walls) without the NACA 0006 airfoil, but including the active grid consisting of nine symmetric airfoil-like vanes, (figure \ref{fig:Numeric_ActiveGrid}). The corresponding inflow conditions are realized by the oscillatory motion of the vanes using a dynamic meshing utility. This utility is based on the so-called `space conservation law', implying  the conservation of the grid cell number during the oscillatory vane motion. Accordingly, an additional equation describing the conservation of space is solved simultaneously with the continuity and momentum equations as well as with equations governing the turbulence quantities. The length of the solution domain accommodating the nine vanes is $2.8$ m (see figure \ref{fig:Numeric_ActiveGrid}), where the zero-gradient outflow boundary conditions were applied. The spanwise extent of the solution domain amounts to 50\% of the NACA 0006 airfoil chord. The fully hexahedral grid consisted of 112,736 cells in the central vertical plane. The three-dimensional computational domain was meshed by ten grid cells covering the spanwise extent leading to the total number of cells corresponding to 1,127,360. 

Whereas the grid resolution at the plenum walls enables the use of exact boundary conditions, the grid resolution at the vane surface is somewhat coarser. Depending on the value of the wall-shear stress, the dimensionless wall distance $y^+$ of the next-to-wall grid node can cover the entire boundary layer span from viscous sublayer and buffer transition zone up to the logarithmic region. Here, the so-called hybrid wall functions proposed originally by \cite{Popovac2007} are adopted and used in conjunction with a differential Reynolds stress model. This wall boundary condition, depending on the value of $y^+$, utilizes a functional blending between the exact boundary conditions corresponding to a fully-resolved boundary layer and the logarithmic law.

Both two-dimensional and three-dimensional computations were performed using the baseline RSM and its eddy-resolving counterpart. The flow conditions corresponding to wake regions generated by the active grid are characterized by intensive bulk flow unsteadiness indicating high turbulent activity. The results obtained by both turbulence models, illustrating the time variation of the streamwise and vertical mean velocity components as well as the gust angle amplitude, are displayed in figure \ref{fig:RSM-IISRSM-Comparison}. The results show no significant difference between two-dimensional and three-dimensional computations. Accordingly, all consequent computations are performed two-dimensionally.

%% file: 04_experimental_considerations.tex
\section{Experimental considerations}

\subsection{Active-grid setup}

Because the parameters $k_2$ and $\alphaG$ are critical for the experimental retrieval of and differentiation between the Sears and Atassi functions, careful attention must be paid to the construction of an apparatus to produce and control gusts with the required character. The flows produced should be as two-dimensional as possible to avoid the influence of three-dimensional unsteady effects on the experimental data. To this end, nine two-dimensional vanes, shown in figure \ref{fig:Setup}, replaced the arrays of diamond-shaped paddles used in the experiments of \cite{cordes_note_2017}. In addition, relations had to be defined so that the gust-angle amplitude $\alphaG$ could be controlled for any given combination of a grid motion frequency and free-stream velocity.

The Sears formulation of the gust-response problem required gusts composed purely of sinusoidal fluctuations in $v$, normal to the airfoil. This implies that in the area above and below the airfoil, fluctuations in $u$ need to be minimized across the height of the wind tunnel. Formulated in the terminology of Atassi's theory, this meant that $k_2$ effectively has to be zero.

The Atassi formulation of the problem, on the other hand, requires these $u$-fluctuations in order to differentiate it from the Sears formulation. This means that $u$ needs to fluctuate along with $v$. For these experiments, the theory of Atassi was interpreted such that the $u$- and $v$-velocities fluctuate with the same frequency, so that both could be controlled by the active grid while keeping $k_1$ and $k_2$ independent from each other.

\subsection{Active-grid protocols}
\label{sec:grid_protocols}
In order to achieve these two gust scenarios using the new active-grid setup, two different grid motion protocols and two different grid configurations were developed. The two motion protocols were defined for the full set of nine vanes. The first protocol, called the focused protocol, involved setting the nine vanes so that their chord lines were all focused to a single point on the test airfoil. As the vanes oscillated, the focal point of the vanes moved across the height of the wind tunnel. This moving focal point directed the flow in a sweep across the height of the tunnel, producing significant fluctuations in $u$ around the test airfoil. This also allowed relatively high gust amplitudes to be maintained for lower grid frequencies and higher free-stream velocities. The second protocol, termed the limited protocol, was simpler, involving basic sinusoidal motions for all vanes. The top and bottom two vanes were however, limited in their motions so that they would not direct flow into the tunnel walls and create unwanted fluctuations. The full range of the vane motions (i.e. twice the amplitude) is defined as $\hat{\theta}$.

In addition to these two grid protocols, the central three vanes of the active grid could be removed, leaving the grid in a six-vane configuration. This construction removed the influence of the wakes of the grid vanes on the test airfoil, which were in part responsible for the fluctuations in $u$ in the nine-vane grid configuration. Similar configurations have been used by \cite{ham_wind_1974} and \cite{lancelot_design_2015} to produce pure sinusoidal fluctuations in $v$ without disturbing the regularity of $u$. This helped minimize fluctuations in $u$, in accordance with the assumptions of the Sears function; thus it was only  used with the limited protocol. Because of this, in combination with the lack of influence from the central three vanes, the gust amplitudes achievable in this configuration were smaller than those from the nine-vane configuration. Conversely, the nine-vane grid configuration was only  used with the focused protocol, so that the fluctuations in $u$ could be maximized, according to the prescriptions of the Atassi function.

Therefore, when the six-vane grid configuration is referenced in this work, it will be assumed that limited protocols were utilized. Similarly, the nine-vane grid configuration will imply that focused protocols were employed.

%% file: 05_measurement_results.tex
\section{Measurement results and discussion}

\subsection{Inflow characterization for the Atassi problem}

In order to verify that distinct inflow conditions for the Sears and Atassi problems could be produced with the active grid configurations detailed above, the flow in the wind tunnel were characterized using hot-wire anemometers and stereo PIV, and additionally with numerical simulations of the wind tunnel.

\begin{figure}
  \centerline{\includegraphics[width=0.5\textwidth]{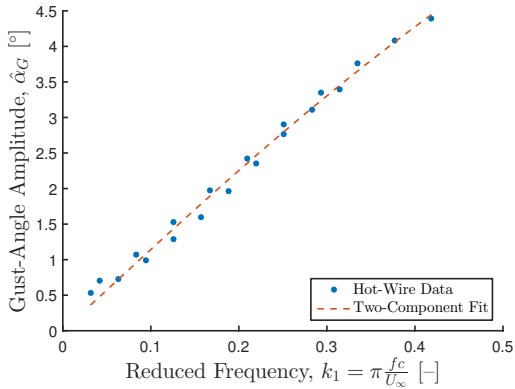}}
  \caption{Two-parameter fit of hot-wire data taken behind the nine-vane active grid configuration at the location of the leading edge of the test airfoil, for ten grid frequencies and three free-stream velocities. The grid vanes moved through an amplitude of $\thetaH = 10^\circ$. Using the known values for $k_1$ and $\alphaG$, values for the Atassi parameters $k_2$ and $\epsilon$ could be estimated using a nonlinear-regression fit based on equation \ref{eqn:AtassiAlphaG}.}
\label{fig:HWfits}
\end{figure}

For each combination of reduced frequencies $k_1$ and grid amplitude $\thetaH$, a gust-angle amplitude $\alphaG$ was computed from the phase-averaged velocity data. An example data set for $\thetaH = 10^\circ$ with the nine-vane grid configuration is shown in figure \ref{fig:HWfits}. For every combination of frequency and free-stream velocity, $\alphaG$ was plotted against $\thetaH$, and a power-law fit was applied to the data. The standard deviation of $\alphaG$ was used to demarcate the region of the parameter space where the grid produces gusts with the quality required by the stringent constraints of the Sears and Atassi theories. Using these fits and limits, given desired values for $\alphaG$ and $k_1$, a corresponding grid amplitude could be selected so that gusts with the right character would be produced.

\begin{figure}
\begin{subfigure}[t]{0.48\textwidth}
\centering
  \includegraphics[width=\textwidth]{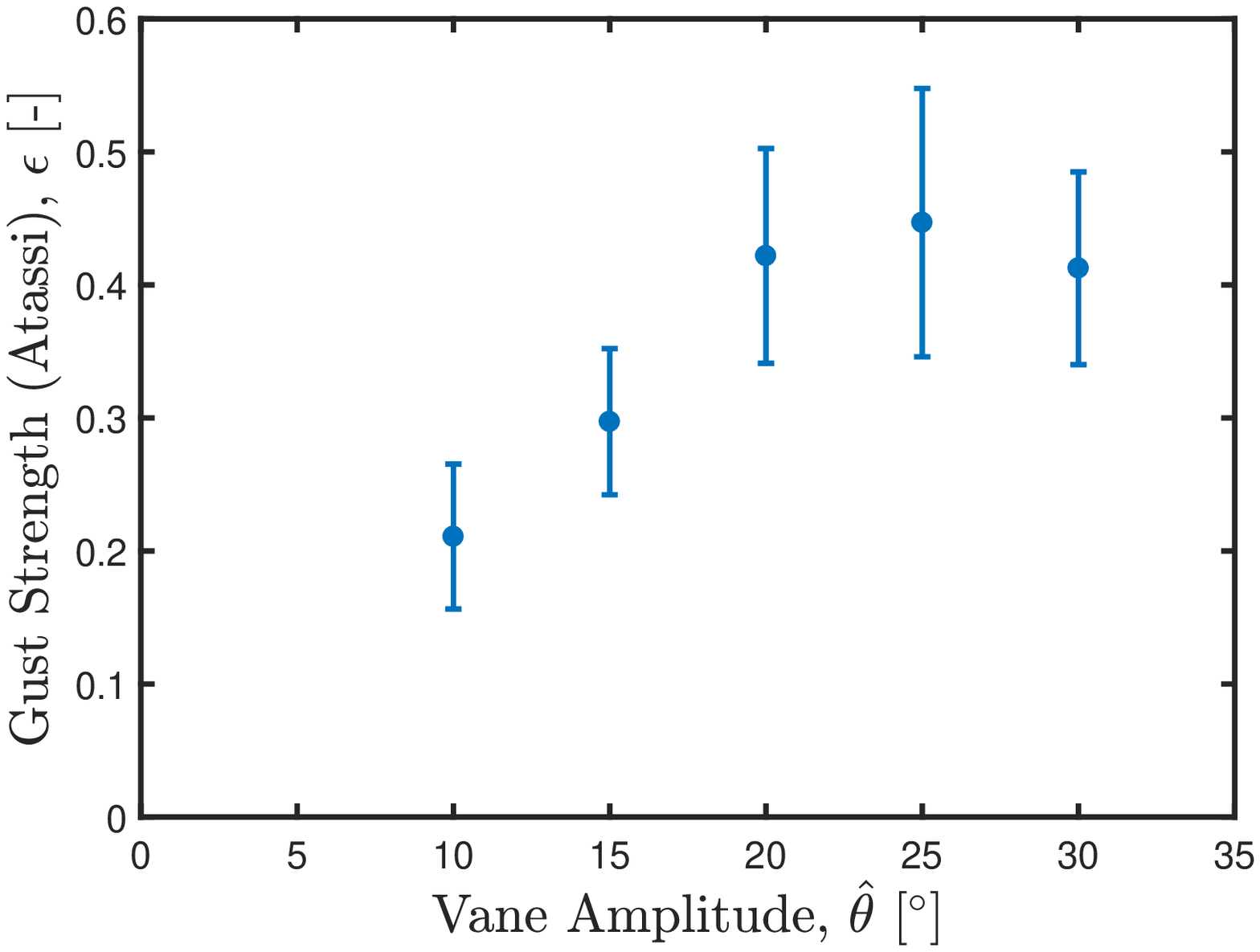}
  \caption{}
\label{fig:HW_eps}
\end{subfigure}
\hfill
\begin{subfigure}[t]{0.48\textwidth}
\centering
  \includegraphics[width=\textwidth]{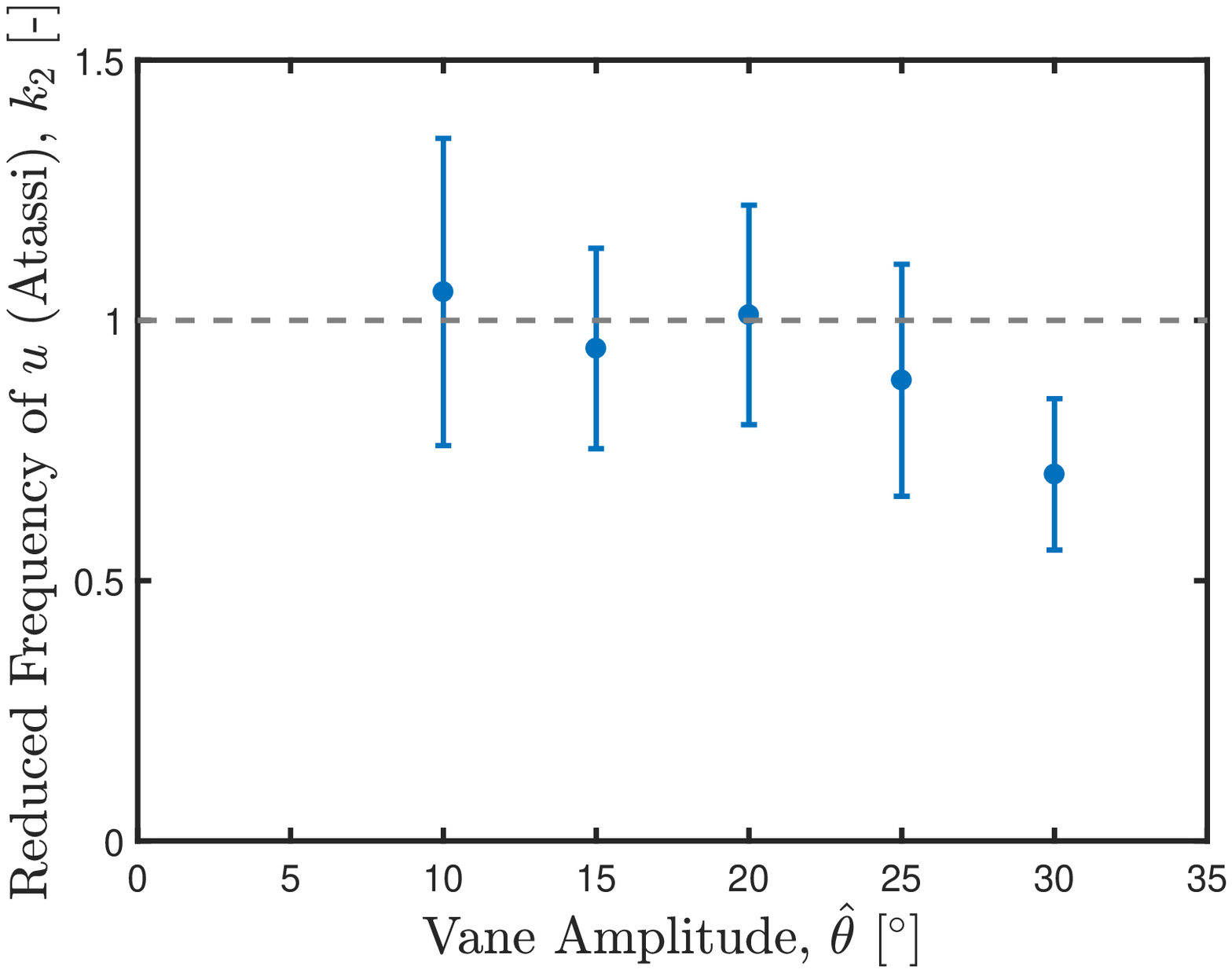}
  \caption{}
\label{fig:HW_k2}
\end{subfigure}
\caption{Estimated values for the parameters $k_2$ and $\epsilon$ for various vane amplitudes in the grid's nine-vane configuration. These were computed using the type of fit shown in figure \ref{fig:HWfits}, and the error bars represent standard errors from the nonlinear-regression coefficients. In the nine-vane configuration, $\epsilon$ increases with vane amplitude until the gust strength saturates, and $k_2$ remains relatively constant across mid-range vane amplitudes. The value of $k_2$ used in these experiments is represented as a dashed line.}
\end{figure}

Additionally, for the data sets taken behind the nine-vane grid configuration, a two-parameter nonlinear-regression fit based on equation \ref{eqn:AtassiAlphaG} was employed to model the data. Through this type of fit, estimates for $\epsilon$ and $k_2$ were computed. As shown in figures \ref{fig:HW_eps} and \ref{fig:HW_k2}, between $\thetaH = 10^\circ$ and $\thetaH = 25^\circ$, $\epsilon$ increased linearly with $\thetaH$, and $k_2$ remained approximately constant within confidence intervals. From $\thetaH = 25^\circ$ on, the effects of full stall on the grid vanes caused these trends to break down. For the nine-vane grid configuration, the grid amplitude could be correlated to the Atassi definition of gust strength as long as stall effects were not considerable. More importantly, the data provided an estimate for $k_2$ -- the deciding factor between the Sears and Atassi theories -- that remained at a constant nonzero value across the range of grid amplitudes employed in the experiments. The average of the estimates below $\thetaH = 25^\circ$, $k_2 = 1.0$, was used to compute Atassi function values for all experiments using the nine-vane grid configuration, since most of the amplitudes used in this study were within the range $5^\circ < \thetaH < 25^\circ$.

\begin{figure}
  \centerline{\includegraphics[width=0.8\textwidth, trim=0 0 0 0 ,clip]{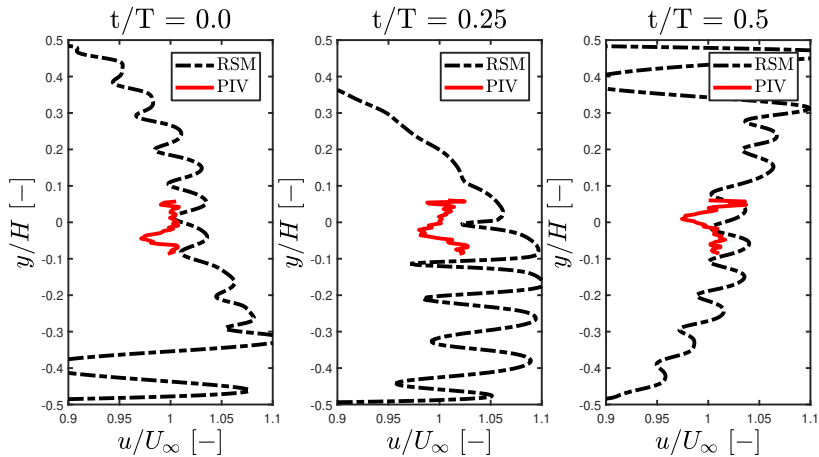}}
  \caption{Variation of the normalized velocity fluctuations in $u$ across the height of the wind tunnel with the active grid running a focused protocol in its nine-vane configuration, as reported by simulations and experiments. The plots represent different dimensionless time instants $t/T$ within one gust period, and $y/H = 0$ represents the location of the leading edge of the airfoil. The experimental data were limited in scope by the width of the laser sheet. Despite some differences in the magnitudes of the $u$ fluctuations, both simulations and experiments show that $u$ does vary significantly in space and time, because of the presence of vanes in the center region of the tunnel as well as the use of focused grid protocols. This implies $k_2$ is nonzero for the nine-vane grid configuration.}
\label{fig:AtassiOnflow_uUinf_yH}
\end{figure}

These results were confirmed in experiments and simulations of the active grid without the test airfoil installed. The $u$ component of the flow velocity is shown to fluctuate significantly in time and across the height of the wind tunnel in figure \ref{fig:AtassiOnflow_uUinf_yH}, which compares PIV data taken in experiments with data from corresponding numerical simulations. The numerical results suggest further that the likely cause of these fluctuations lies in the combination of the wake profiles of the vanes in the center region of the tunnel and the use of focused protocols, which are responsible for the large-scale undulations in velocity across the height of the tunnel. These results confirmed that the value of $k_2$ for the Atassi function is significantly greater than zero for the nine-vane configuration.

\subsection{Inflow characterization for the Sears problem}

\begin{figure}
  \centerline{\includegraphics[width=0.8\textwidth, trim=0 0 0 0 ,clip]{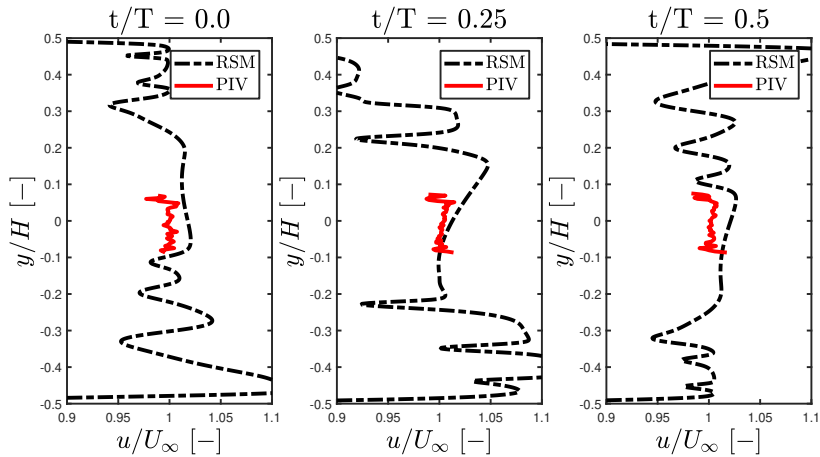}}
  \caption{Variation of the normalized velocity fluctuations in $u$ across the height of the wind tunnel with the active grid running a limited protocol in its six-vane configuration, as reported by simulations and experiments. Three dimensionless time instants $t/T$ within one gust period are given. In comparison with the results from the nine-vane configuration shown in figure \ref{fig:AtassiOnflow_uUinf_yH}, both simulations and experiments show that $u$ remains effectively constant in the region of the test airfoil. This implies $k_2 = 0$ for the six-vane grid configuration.}
\label{fig:SearsOnflow_uUinf_yH}
\end{figure}

\begin{figure}
  \centerline{\includegraphics[width=0.8\textwidth, trim=0 0 0 0 ,clip]{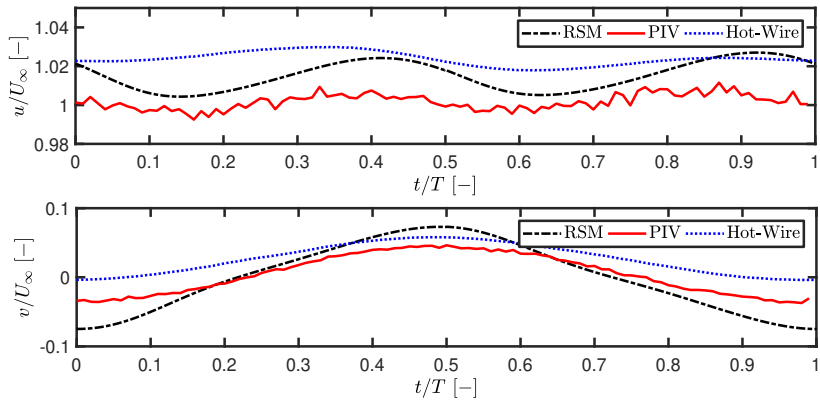}}
  \caption{Evolution of the normalized velocity fluctuations in $u$ and $v$ over a single gust period with the six-vane configuration and limited protocols, as reported by simulations and experiments (PIV and hot-wire probe data). In both trend and magnitude, the simulations and experiments show remarkable agreement for both velocity components. As in figure \ref{fig:SearsOnflow_uUinf_yH}, it is clear that the variations in $u$ are very small, implying $k_2$ is effectively zero in the six-vane configuration.}
\label{fig:SearsOnflow_tT_uUinf}
\end{figure}

The hot-wire data were also used to determine relationships between $\alphaG$ and $\thetaH$ for the six-vane grid configuration. Since the goal of using this grid configuration was to achieve $k_2 \approx 0$, it was not necessary to estimate $k_2$ and $\epsilon$ using the two-parameter fits discussed above. Thus, the fluctuations in the $u$-component of the flow velocity were observed as an indicator of the presence of a significant $k_2$-fluctuation in the same manner as described in the section above. To this end, results from both stereo PIV experiments in the empty tunnel and numerical simulations of the setup were compared. Figures \ref{fig:SearsOnflow_uUinf_yH} and \ref{fig:SearsOnflow_tT_uUinf} show the experimental and simulated properties of the flow field from grid motions at $f = 10$ Hz, $U_\infty = 15$ m/s, and $\thetaH = 11.73^\circ$. The $u$-profile across the height of the tunnel, shown in figure \ref{fig:SearsOnflow_uUinf_yH}, remained relatively uniform across the central region of the tunnel in both simulations and experiments, with no evidence of the temporal and spatial fluctuations present in figure \ref{fig:AtassiOnflow_uUinf_yH} that signified a nonzero $k_2$. Additionally, the fluctuations of $u$ in time were very small (on the order of 1\% of $U_\infty$), further suggesting that $k_2$ for the six-vane grid configuration was effectively zero. The fact that the PIV data and data from the numerical simulations exhibited such strong agreement suggests that the simulations can be used reliably to characterize the inflow conditions in the tunnel in the region of the test airfoil.

\subsection{Measurements of the Sears problem}

\begin{figure}
  \centerline{\includegraphics[width=0.5\textwidth]{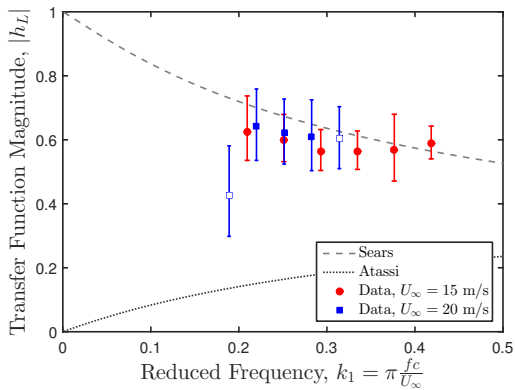}}
  \caption{Gust response of a NACA-0006 airfoil under Sears inflow conditions for a variation of the reduced frequency $k_{1}$ and a gust-angle amplitude of $\alphaG = 2^\circ$. Aside from the single outlier, the data follow the general trend of the Sears function.}
\label{fig:SearsNACAAlphaGConst}
\end{figure}

After the flows produced by the active grid had been characterised, the two grid configurations were used to investigate the Sears and Atassi functions in their respective flow situations. The Sears function was tested using the six-vane grid configuration and limited grid protocols. Tests were conducted using two free-stream velocities of $U_\infty = 15$ and $20$ m/s (corresponding to Reynolds numbers of $200,000$ and $260,000$) over a range of gust frequencies, with vane amplitudes tuned to produce gusts with $\alphaG = 2^\circ$. This amplitude was chosen to avoid dynamic effects and flow separation on the thin airfoil while maintaining a high enough signal-to-noise ratio from the force balances. The experimental data from this investigation are shown in figure \ref{fig:SearsNACAAlphaGConst}. These data follow the trend of the Sears function reasonably well within one standard error, apart from one outlier, which the enlarged error bars would suggest was likely due to experimental errors in achieving the desired $\alphaG$ for this particular case.

\begin{figure}
  \centerline{\includegraphics[width=0.5\textwidth]{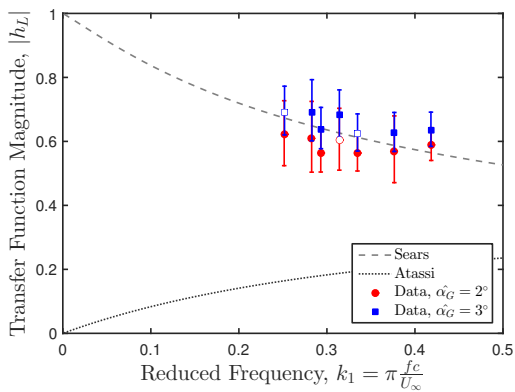}}
  \caption{Gust response of a NACA-0006 airfoil under Sears inflow conditions for a variation of the reduced frequency $k_{1}$. Two gust-angle amplitudes ($\alphaG = 2^\circ$ and $\alphaG = 3^\circ$) are shown. As predicted by the theory, there is little difference between the two cases. The slight offset in the $3^\circ$ case could be due to the onset of dynamic effects on the thin airfoil.}
\label{fig:SearsNACAAlphaGVar}
\end{figure}

To verify that $\alphaG$ does not affect experimental correspondence with the Sears function, a series of gusts with $\alphaG = 3^\circ$ was also tested. These data, shown in figure \ref{fig:SearsNACAAlphaGVar}, lie slightly above the $\alphaG = 2^\circ$ data, likely because of the dynamic effects that would have enhanced the amplitude of the lift fluctuations at this relatively high angle of attack. However, the data still follow the Sears trend within one standard error, thus confirming that the shape of the Sears function holds irrespective of gust strength, as long as flow separation does not occur. Values of $\alphaG$ smaller than $2^\circ$ were also tested, but these did not exhibit a clear trend due to the influence of noise. 

\subsection{Measurements of the Atassi problem}

While the gust strength in the Sears problem is represented solely by $\alphaG$, the analogous parameter in the Atassi formulation is $\epsilon$, which, when $k_2$ is nonzero, is a nontrivial function of $\alphaG$. To show that the Atassi function, like the Sears function, does not depend on the gust strength, $\epsilon$ can be held constant. According to equation \ref{eqn:AtassiAlphaG}, a constant $\epsilon$ means that $\alphaG$ will increase with $k_1$. This could suggest, however, that the primary difference between the Sears and Atassi functions is the behavior of $\alphaG$. Thus, a more direct comparison with the results obtained for the Sears function can be made if $\alphaG$ is held constant. Since $\alphaG$ is a function of $\epsilon$, neither parameter should affect the shape of the Atassi function, thus leaving $k_2$ as the sole differentiating factor between the Sears and Atassi theories.

Either one of these parameters can be held constant over a given range of reduced frequencies by setting the vane amplitude $\thetaH$ of the active grid accordingly. As outlined for the six-vane grid configuration used to validate the Sears function, changing $\thetaH$ allows $\alphaG$ to be controlled for a given value of $k_1$. $\epsilon$, on the other hand, scales proportionally with $\thetaH$ without depending on $k_1$ (cf.\ figure \ref{fig:HW_eps}), thus implying that a constant $\thetaH$ yields a constant value of $\epsilon$ irrespective of $k_1$ and $\alphaG$. The latter variant was employed by \cite{cordes_note_2017} using the older, three-dimensional configuration of the Oldenburg active grid.

\begin{figure}
\begin{subfigure}[t]{0.48\textwidth}
\centering
  \includegraphics[width=\textwidth]{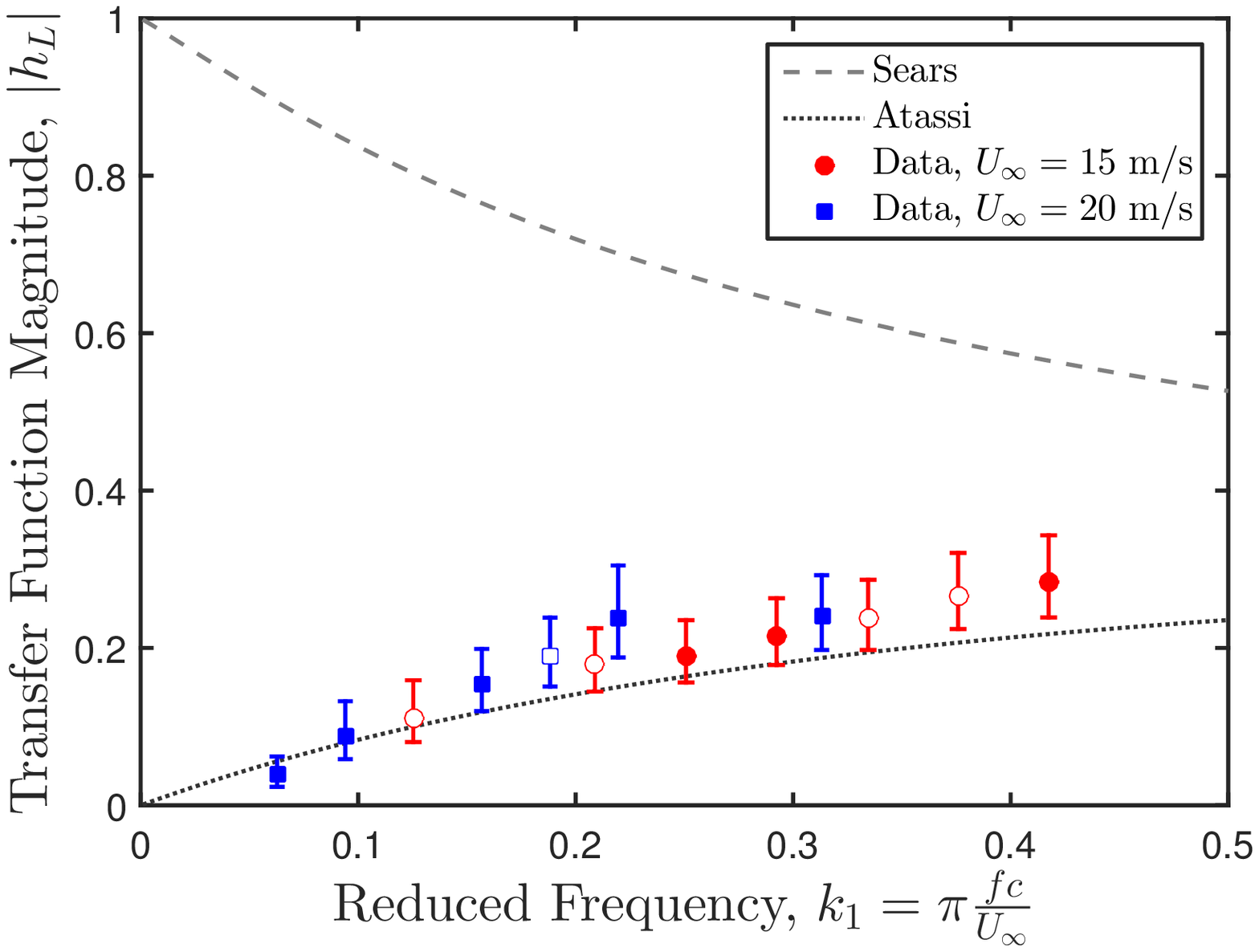}
  \caption{}
\label{fig:AtassiNACAThetaConst}
\end{subfigure}
\hfill
\begin{subfigure}[t]{0.48\textwidth}
\centering
  \includegraphics[width=\textwidth]{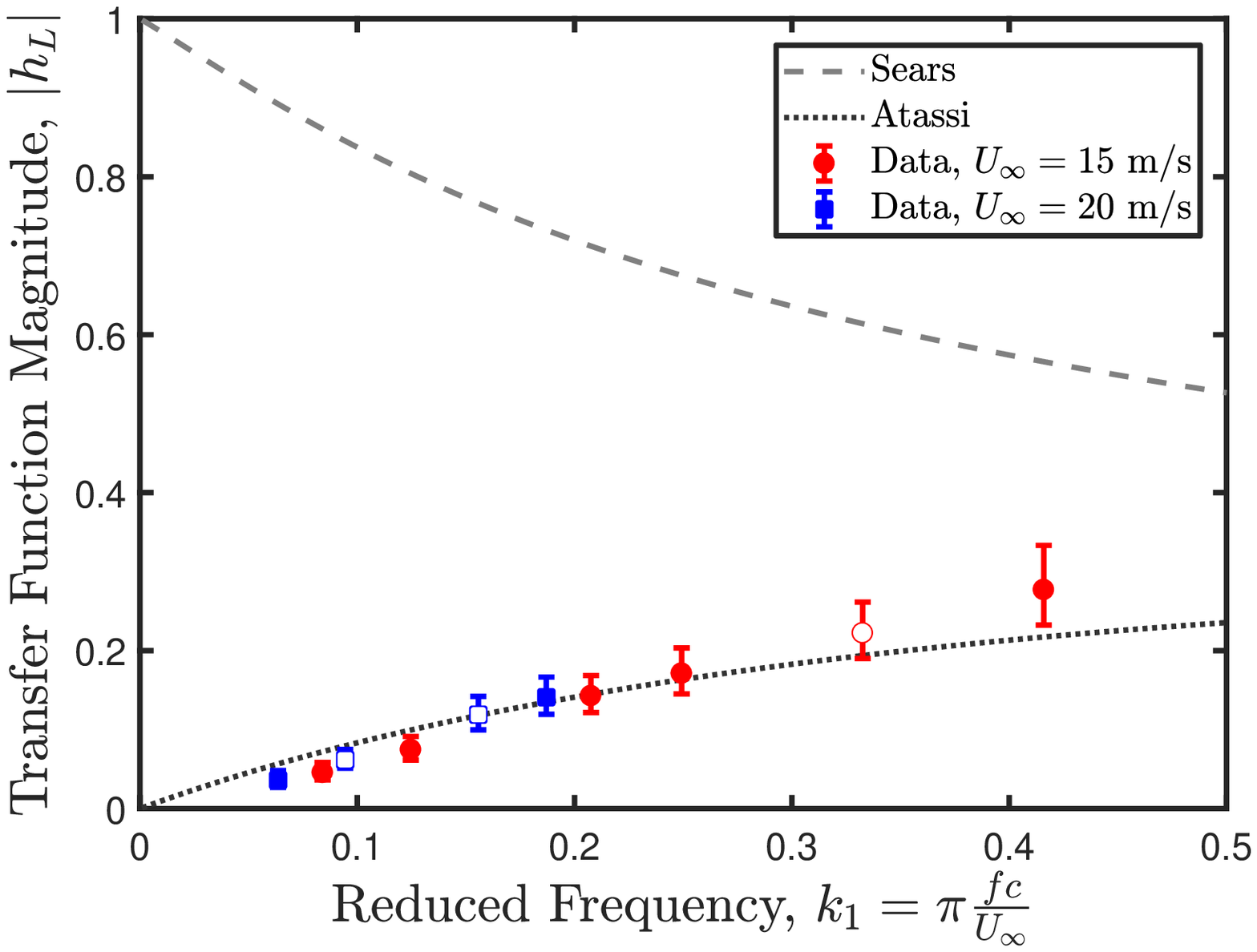}
  \caption{}
\label{fig:AtassiNACAAlphaGConst}
\end{subfigure}
\caption{Gust response of a NACA-0006 airfoil under Atassi inflow conditions for a variation of the reduced frequency $k_{1}$. In the left plot, the flap amplitude is kept constant at $\thetaH = 5^\circ$, while in the right plot, the gust-angle amplitude is kept constant at $\alphaG = 2^\circ$. Both variations show clear correspondence with the Atassi function, and demonstrate that the function is independent from both $\epsilon$ and $\alphaG$.}
\end{figure}

These variants were tested in experiments using the nine-vane grid configuration. First, the gust strength according to the Atassi formulation was held constant using a constant grid amplitude of $\thetaH = 5^\circ$. Then, another test series was conducted using a constant gust-angle amplitude of $\alphaG = 2^\circ$. Due to the limitations of the active grid and the gusts it could produce, the reduced frequency range of the second series was smaller than that of the first. The value of $k_2$ was fixed at $k_2 = 1$, taken from the analysis of the hot-wire probe data detailed previously. The use of the NACA-0006 airfoil at zero mean angle of attack allowed the correction terms $\alpha_\beta$ and $\alpha_m$ of the Atassi function to be set to zero. The results from these experiments are shown in figures \ref{fig:AtassiNACAThetaConst} and \ref{fig:AtassiNACAAlphaGConst}, respectively. Both series show a close correspondence with the predictions of the Atassi function. This confirms experimentally that the difference between the theories of Sears and Atassi lies exclusively in the value of $k_2$ and not in the character of $\alphaG$.

\begin{figure}
\begin{subfigure}[t]{0.48\textwidth}
\centering
  \includegraphics[width=\textwidth]{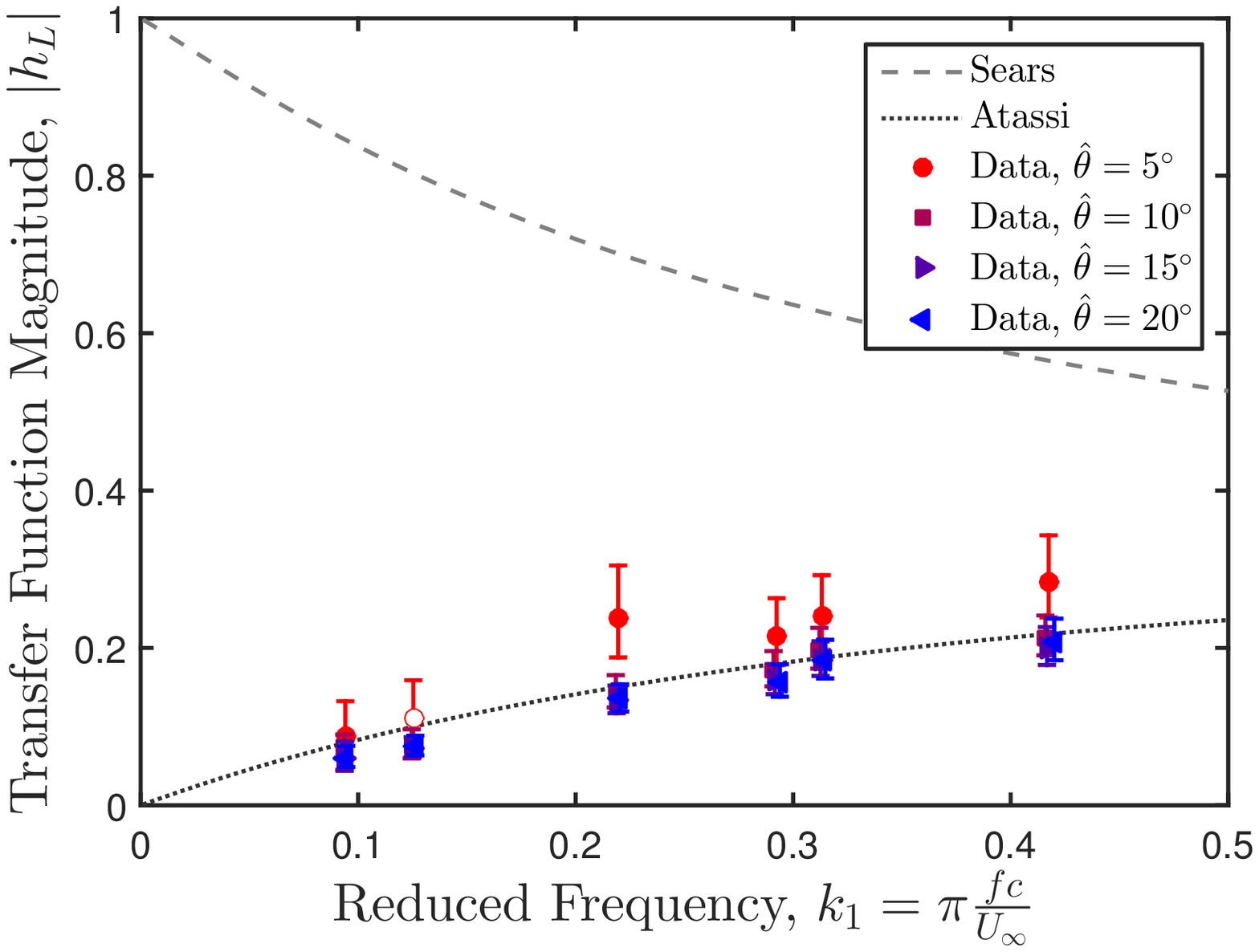}
  \caption{}
\label{fig:AtassiNACAThetaVar}
\end{subfigure}
\hfill
\begin{subfigure}[t]{0.48\textwidth}
\centering
  \includegraphics[width=\textwidth]{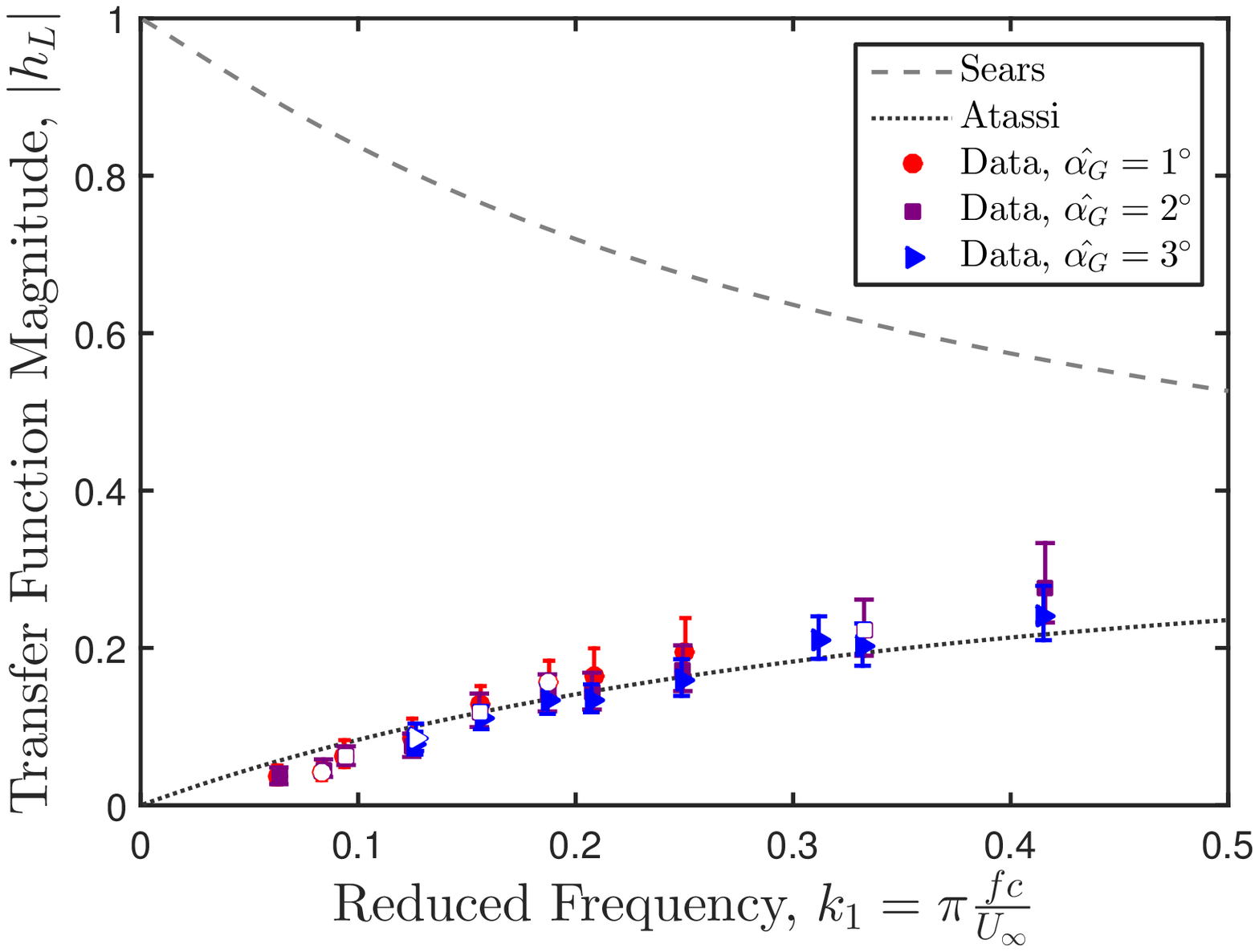}
  \caption{}
\label{fig:AtassiNACAAlphaGVar}
\end{subfigure}
\caption{Gust response of a NACA-0006 airfoil under Atassi inflow conditions for a variation of the reduced frequency $k_{1}$, while the streamwise reduced frequency was $k_2 = 1$. On the left, the flap amplitude $\thetaH$ was varied, and on the right the gust-angle amplitude $\alphaG$ was varied. In both cases, the data follow the trend of the Atassi function, demonstrating further that the function is independent from gust strength.}
\end{figure}

The tests were extended to a range of grid amplitudes and gust-angle amplitudes, in order to demonstrate that the Atassi function is independent of $\alphaG$ and $\epsilon$. The results for $5^\circ \leq \thetaH \leq 20^\circ$ and $1^\circ \leq \alphaG \leq 3^\circ$ are shown in figures \ref{fig:AtassiNACAThetaVar} and \ref{fig:AtassiNACAAlphaGVar}, respectively. As expected, the trend of the data remained unchanged, dictated only by the value of $k_2$ of the setup. This confirmed that the normalization used for the Atassi-function data was correct and worked in practice to collapse the data.

Two differences between these results and those from the Sears function experiments call for a brief elucidation. First, a higher percentage of data points were marked by the force-data validation criterion in the Atassi function data than in the Sears function data. This is due to the influence of the wakes of the central vanes, which are not present in the six-vane grid configuration and add higher frequency fluctuations to the sinusoidal gust profiles. These tagged points exhibited neither larger error bars nor significant deviation from the predictions of the Atassi function, so the influence of these higher frequency superpositions was deemed negligible for the purposes of this study. Second, the error bars of the Atassi function data appear to be significantly smaller than those of the Sears function data. The error bars, however, are computed relative to the value of the transfer function; thus, for smaller transfer function magnitudes, the error bars will be proportionally smaller.

These two points notwithstanding, the data from the Atassi function experiments still seem to converge to the trend of the Atassi function better than the Sears function data converge to the Sears function trend. This observation will be treated in the next section. Two major conclusions can  be drawn from the experimental validations of the Sears and Atassi functions presented here. First, the experimental data confirm that the two theories can be measured in real flows according to their fundamental assumptions. Specifically, the normalization factors for both theories are appropriate and effectively remove the influence of the gust-strength parameters $\alphaG$ and $\epsilon$, allowing the experimental data to collapse onto the theoretical predictions. Second, the results confirm experimentally that the difference between the Sears and Atassi functions lies not in the Atassi corrections for mean angle of attack or camber, but rather solely in the character of the streamwise gusts as captured by the $k_2$ parameter.

\subsection{Influence of inflow turbulence level}

It has been noted that the convergence of the data around the Atassi function appears to be better than that observed in the Sears function experiments. Part of this can be explained by the relative sizes of the error bars between these two data sets. However, the presence of the central vanes in the nine-vane grid configuration used for the Atassi function experiments suggests a more physical interpretation, one that should be taken into account for a more balanced comparison of the theories of Sears and Atassi.

\begin{figure}
  \centerline{\includegraphics[width=0.5\textwidth]{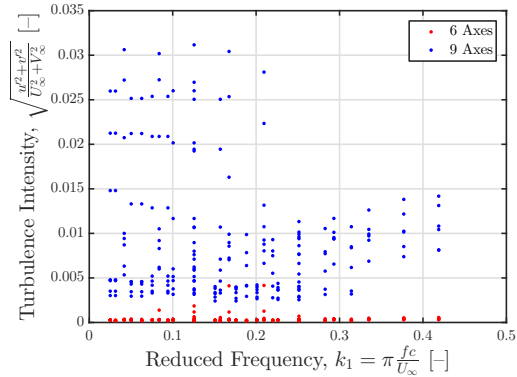}}
  \caption{Turbulence intensities of the inflow generated by the six- and nine-vane grid configurations, computed from hot-wire measurements at the position of the leading edge of the test airfoil. The sinusoidal velocity fluctuation at the grid's motion frequency was first filtered out so that the intensities of the background turbulence could be compared. The results are plotted against the reduced frequency $k_{1}$ to show that the turbulence intensities behind the nine-vane grid were consistently higher than those behind the six-vane grid. This suggests that the better convergence of the Atassi condition data is in part due to the higher background turbulence intensities in the nine-vane grid configuration.
  }
\label{fig:TurbulenceLevelInflow}
\end{figure}

\cite{jancauskas_aerodynamic_1983} showed that increasing the background turbulence intensity in gusty flows on bridge-deck sections leads to better correspondence with the Sears function. In the nine-vane grid configuration, the wakes of the central vanes were responsible for higher turbulence intensities in the flows impinging on the airfoil than in the six-vane grid configuration where the middle section of the tunnel was free from disturbances. This was determined using the hot-wire probe data from the tunnel characterisation experiments. A centered moving average over 100 data samples was applied to each velocity component in order to extract the turbulent velocity fluctuations from the generally sinusoidal velocity profile of the gust. A turbulence intensity was then computed from these two velocity fluctuation components for every combination of $k_1$ and $\thetaH$ tested. The distribution of these turbulence intensity values is represented in figure \ref{fig:TurbulenceLevelInflow}. From these data, it is clear that the central three axes in the nine-vane grid configuration generated uniformly higher background turbulence levels than were observed in the six-vane grid configuration.

This observation implies that the perceived better convergence seen in the Atassi function data could be due in part to the higher levels of background turbulence present in these experiments. To test this hypothesis, higher turbulence flow conditions were generated on the test airfoil for the Sears function experimental setup. This could not simply be done by replacing the vanes in the middle of the tunnel, as this would have made $k_2$ nonzero. Furthermore, unlike the apparatus of Jancauskas and Melbourne, turbulence generating screens could not  be easily mounted in the tunnel. Therefore, turbulence was generated directly on the test airfoil using a thin strip of tape, intended to trip transition to turbulence in the boundary layer. The strips of tripping tape used were 3.5 mm wide and 0.1 mm thick, and were placed symmetrically on both sides of the airfoil. Three separate locations along the chord of the airfoil were tested: $x = 0.05c$, $0.10c$, and $0.40c$. The set of reduced frequencies and free-stream velocities from the original Sears function series was used for these three tape locations.

\begin{figure}
	\begin{subfigure}[t]{0.48\textwidth}
		\centering
		\includegraphics[width=\textwidth]{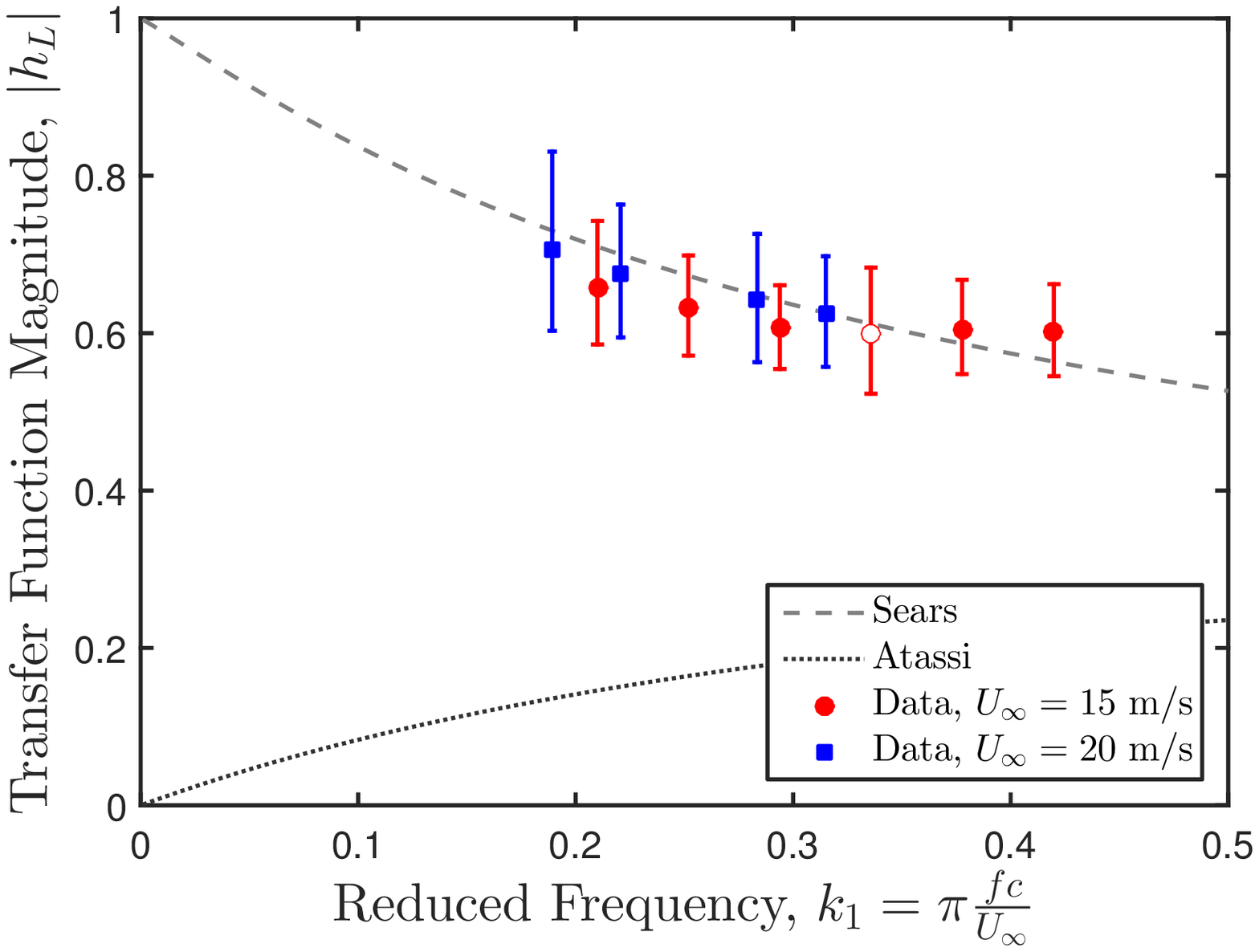}
		\caption{}
		\label{fig:SearsTrip0}
	\end{subfigure}
	\hfill
	\begin{subfigure}[t]{0.48\textwidth}
		\centering
		\includegraphics[width=\textwidth]{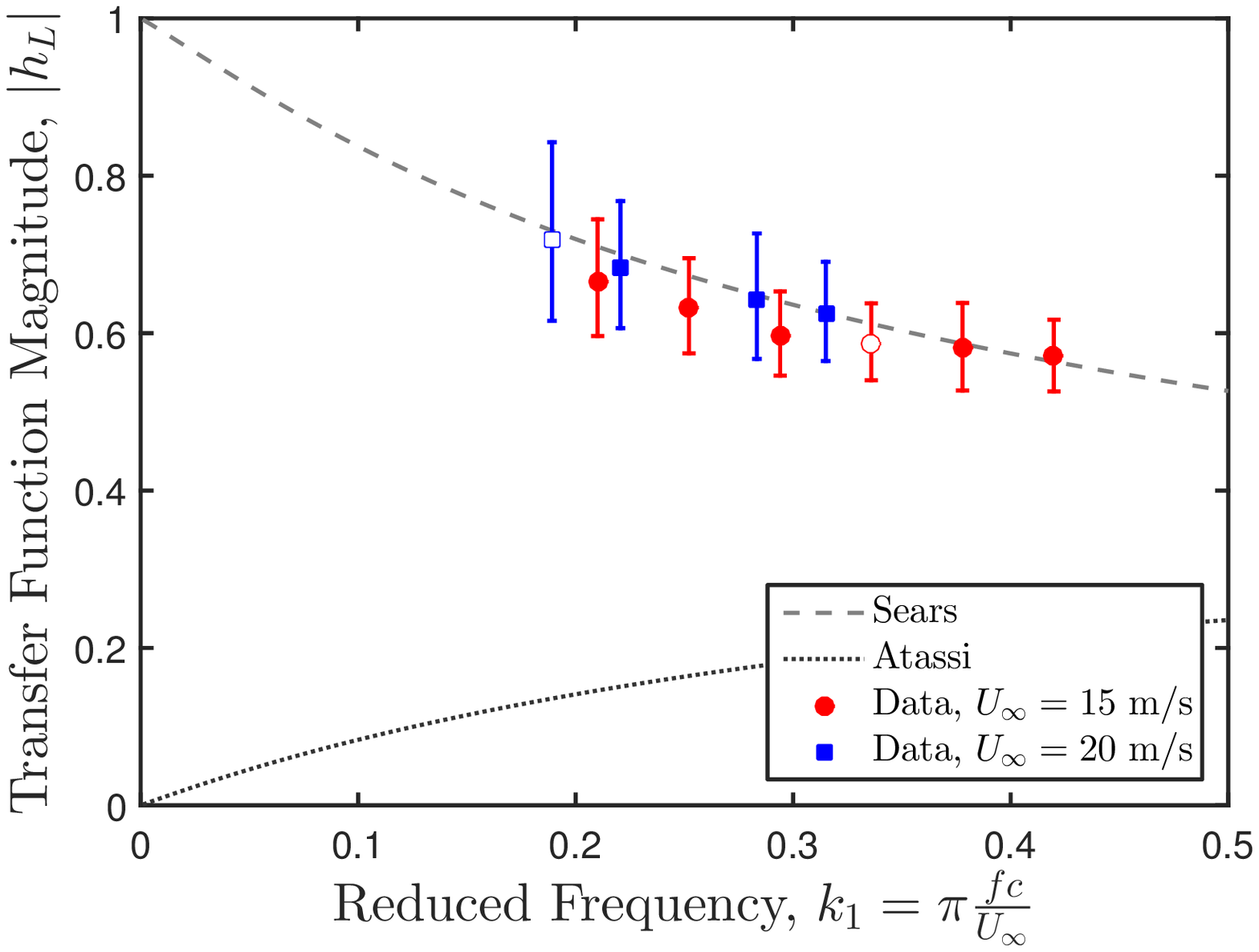}
		\caption{}
		\label{fig:SearsTrip10}
	\end{subfigure}
	\centering
	\begin{subfigure}[t]{0.48\textwidth}
		\centering
		\includegraphics[width=\textwidth]{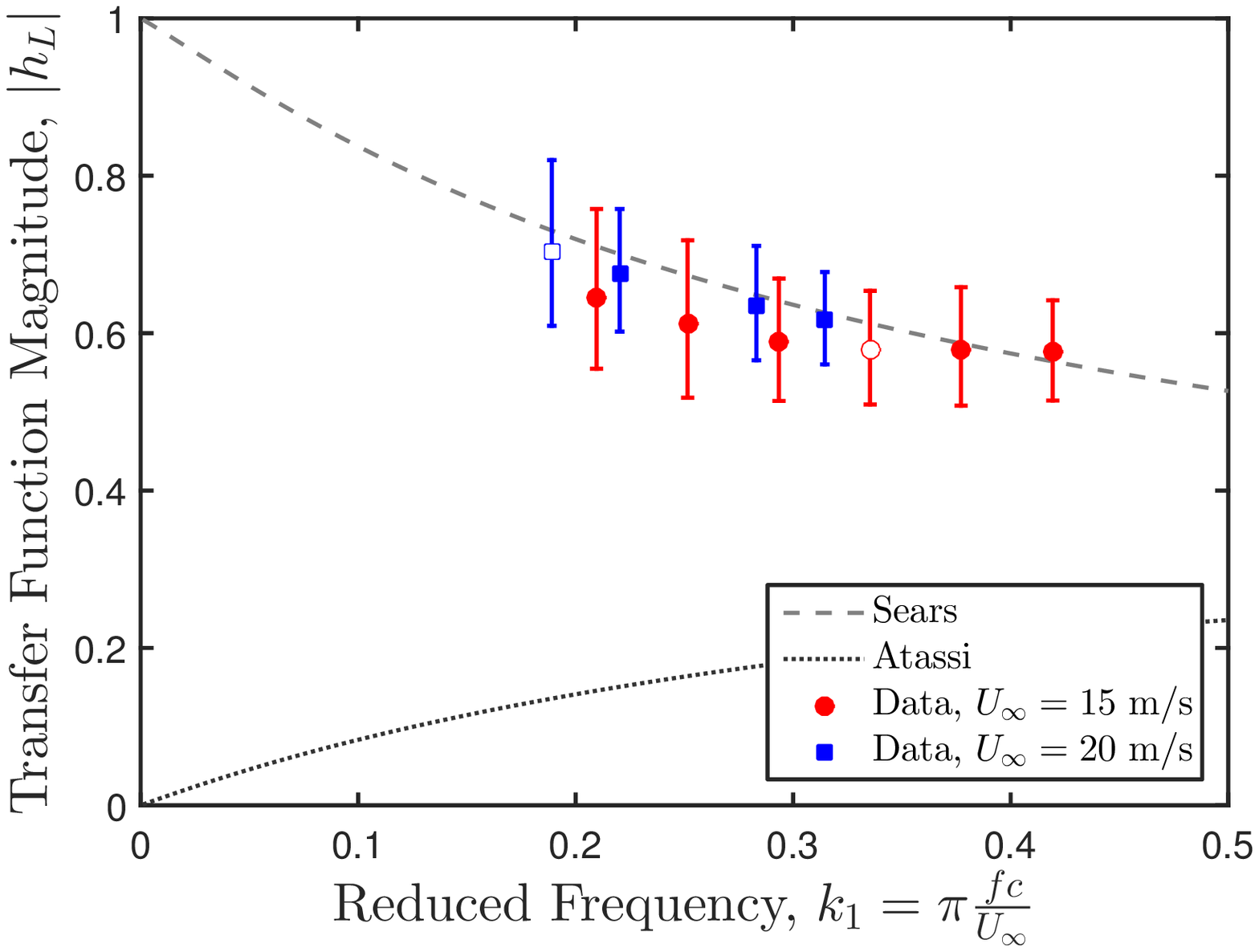}
		\caption{}
		\label{fig:SearsTrip40}
	\end{subfigure}
	\caption{Gust response of a NACA-0006 airfoil under Sears inflow conditions for a variation of the reduced frequency $k_{1}$, with the addition of tripping tape at (a) 5\%, (b) 10\%, and (c) 40\% of the chord length behind the leading edge. The gust-angle amplitude was held constant at $\alphaG = 2^\circ$. The convergence to the Sears function was significantly improved from the clean airfoil configuration: the mean square error values of (a) $8.79\times10^{-4}$, (b) $6.62\times10^{-4}$, and (c) $1.36\times10^{-3}$ were all lower than that of the unmodified airfoil ($1.12\times10^{-2}$). The tape location nearest to the airfoil's point of highest curvature corresponded to the best correspondence with the Sears function.}
\end{figure}

The data from these series of experiments are shown in figures \ref{fig:SearsTrip0}, \ref{fig:SearsTrip10}, and \ref{fig:SearsTrip40}. In all three cases, the correspondence with the Sears function is significantly improved compared to the case without tape given in figure \ref{fig:SearsNACAAlphaGConst}. This result itself confirms in retrospect that the tripping tape did indeed affect the character of the boundary layer, as all other experimental parameters remained the same between the series. The degree of correspondence can be quantified by comparing the mean square error between each data set and the Sears function: for the tape locations of $x = 0.05c$, $0.10c$, and $0.40c$, the mean square errors were $8.79\times10^{-4}$, $6.62\times10^{-4}$, and $1.36\times10^{-3}$. These values are at least one order of magnitude lower than the case without tripping tape, which had a mean square error of $1.12\times10^{-2}$. The best convergence was observed in the case where the tripping tape was located nearest to the point of greatest curvature on the airfoil.

Jancauskas and Melbourne speculated that the physical cause of the improved convergence with increasing turbulence intensity lay in the fact that turbulent flows generally follow the shapes of profiles better than laminar flows. According to this line of reasoning, flows with high levels of background turbulence would reduce separation and boundary-layer thickness effects, thereby making the overall flow field fit the assumptions of Sears more closely. The ambiguity with their experiments, however, was that the incoming gusts themselves were turbulent, and therefore the root cause of the effect could not be localized to the surface of the test profile itself. In this study, the gusts produced had very low levels of background turbulence, and therefore the cause of the improved convergence could be isolated to the surface of the airfoil. The character of the boundary layer on the airfoil is thus a critical parameter for determining how well the Sears function matches experimental data. The mechanism by which this occurs remains unclear. The hypothesis of Jancauskas and Melbourne is still a feasible explanation; another explanation could stem from the argument that the transport of momentum from the gust to the airfoil is more effective in turbulent rather than laminar boundary layers. The latter hypothesis would explain why no clear trend was found when gusts with $\alphaG = 1^\circ$ were used. In either case, the results presented here demonstrate that the presence of turbulent boundary layers on an airfoil leads to better correspondence with the Sears function. This finding is rather counterintuitive in light of the potential-flow assumptions of the Sears theory, and suggest that the Sears function may be far more applicable in real-world flow situations than its first-order origins would make it appear. This particular hypothesis will be addressed in future studies.

\subsection{Normalization}

The importance of the $k_2$ parameter in differentiating between the theories of Sears and Atassi has thus been demonstrated in a controlled experimental setting, including investigations using various gust strengths and different levels of turbulence on the airfoil. In order to complete this analysis, it is instructive to determine the extent to which the simpler Sears function applies when $k_2$ is nonzero. This serves to quantify the degree of difference between the Sears and Atassi functions in practice, as well as to determine the relative significance of the streamwise gusts represented by $k_2$ compared to the normal gusts given by $k_1$.

\begin{figure}
\begin{subfigure}[t]{0.48\textwidth}
  \centering
  \includegraphics[width=\textwidth]{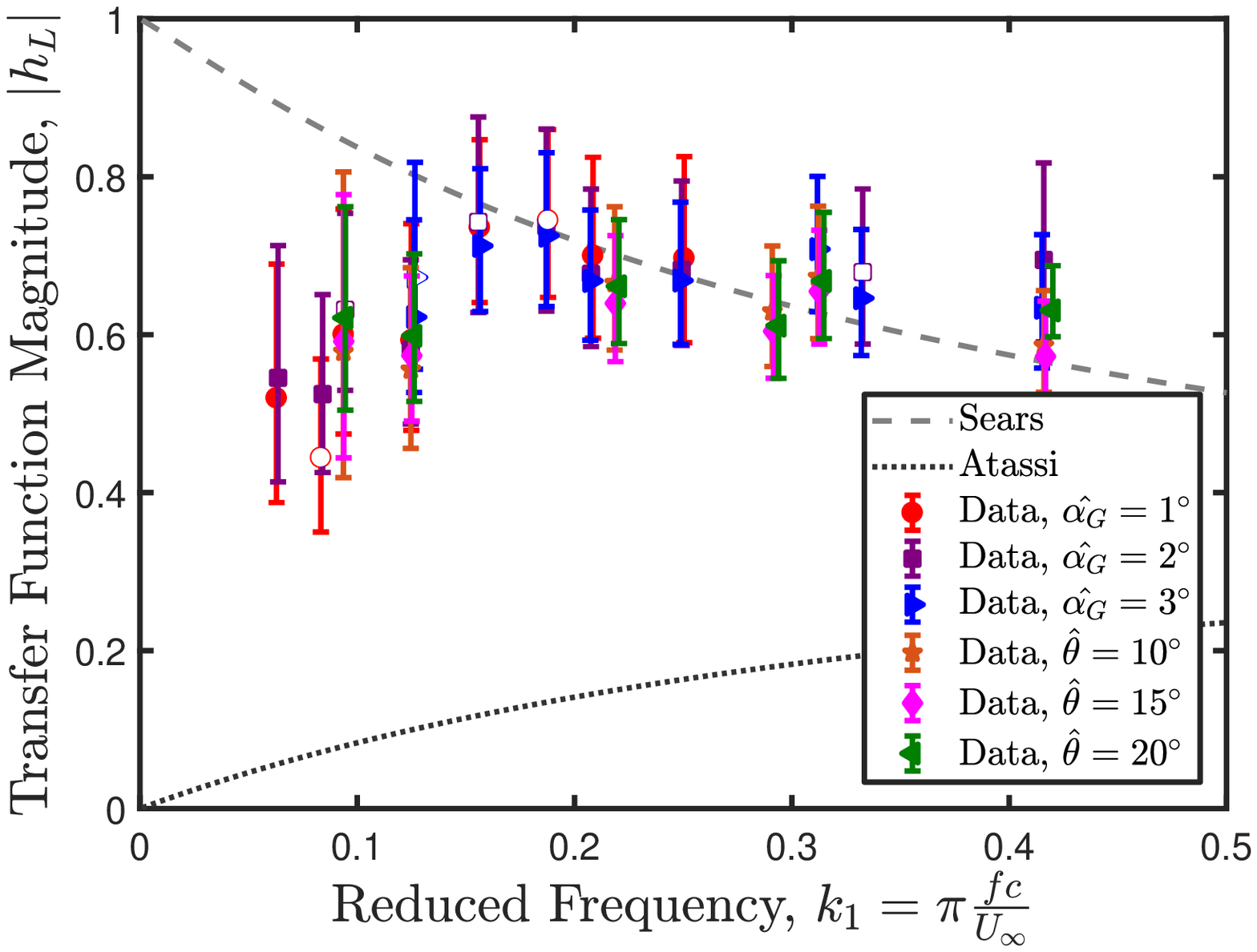}
  \caption{}
\label{fig:AtassiWSearsNorm}
\end{subfigure}
\hfill
\begin{subfigure}[t]{0.48\textwidth}
  \centering
  \includegraphics[width=\textwidth]{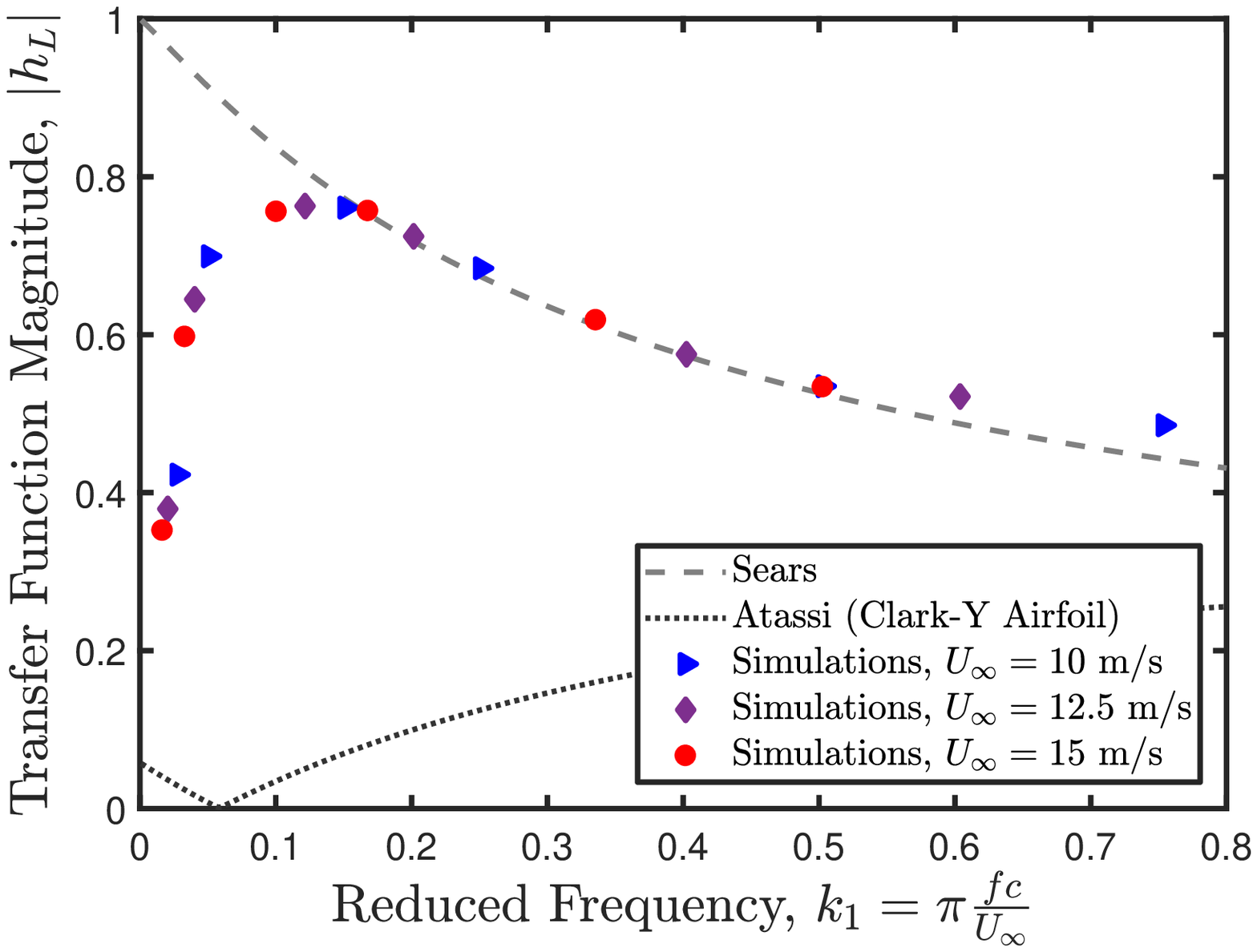}
  \caption{}
\label{fig:AtassiWSearsNormSim}
\end{subfigure}
\caption{The data shown in figure \ref{fig:AtassiWSearsNorm} are taken from figures \ref{fig:AtassiNACAThetaConst} and \ref{fig:AtassiNACAAlphaGConst} (nine-vane grid configuration) and are instead normalized using the Sears normalization. These are compared with 2D RSM simulations of the setup of \cite{cordes_note_2017} with a Clark-Y airfoil (figure \ref{fig:AtassiWSearsNormSim}), also using a nine-vane configuration and applying the Sears normalization to the data. Both the experimental and numerical data show that, for higher reduced frequencies, the Sears function can be applied even when Atassi conditions are present.}
\end{figure}

To address these points, data taken under Atassi conditions were simply normalized using the Sears normalization factor. This approach effectively ignored the streamwise gusts present in the experiment and assumed the Sears function could be applied to the situation. Since the Atassi normalization had been shown to be effective, the data shown in figures \ref{fig:AtassiNACAThetaConst} and \ref{fig:AtassiNACAAlphaGConst} were all compiled in a single plot and normalized by the Sears normalization factor. This combined data set is shown in figure \ref{fig:AtassiWSearsNorm}. Though three values each of $\alphaG$ and $\thetaH$ were used, the data followed a clear trend, indicating that the Sears function remained independent of $\alpha_G$ and was not a function of $\epsilon$. This trend exhibited agreement with the Sears function for reduced frequencies above $k_1 \approx 0.15$, below which significantly lower transfer function magnitudes were observed. This trend implies that for a fixed $k_2$, as $k_1$ increases the streamwise gusts become less significant to the formulation of the gust-response problem, and thus the Sears function is better able to model the unsteady lift forces. Because only a single value of $k_2$ was tested in these experiments, it was not possible to determine a more general relation for the critical value of $k_1$ at which the Sears function becomes applicable.

Numerical simulations of a similar problem confirmed this finding in a different experimental context. These simulations were carried out to model the experiment of \cite{cordes_note_2017}, which had obtained a general Atassi trend despite the use of an incorrect normalization. A Clark-Y profile (11.7\% thickness) with a chord length of $c = 0.18$ m and at zero angle of attack was used as the test airfoil, and 2D RSM simulations were carried out at three free-stream velocities and a range of grid frequencies. The grid was modeled with focused protocols and amplitudes of $\thetaH = 30^\circ$, thus producing gusts with a non-negligible value of $k_2$. The forces on the airfoil were computed from the flow field. More information regarding the setup of these simulations is given in \cite{wegt_numerische_2017}, and the resulting transfer-function values -- again normalized according to the Sears normalization -- are shown in figure \ref{fig:AtassiWSearsNormSim}. In spite of the use of different airfoil profiles, the same trend as that observed in the experiments detailed above was retrieved in the numerical data. An exact value of $k_2$ was not available from the data, so the specific reduced frequency where the data converged to the Sears function could not be directly compared with that from the experimental data. Nevertheless, the numerical results confirmed the result from experiments, that the influence of the $k_2$ gust decreases with increasing $k_1$.

The results of these experiments demonstrate that for a given value of $k_2$, there exists little difference between the Sears and Atassi functions at high values of $k_1$. This result was suggested by \cite{atassi_sears_1984} in a limiting case analysis for $k_1 \rightarrow \infty$, so the data shown here simply confirm this mathematical result. This observation, however, offers an explanation for the trends seen in previous experiments involving gusts generated by a turbulence grid \cite[e.g.][]{larose_experimental_1999, hatanaka_new_2002, lysak_measurement_2016}. The gusts in these studies would have had significant $k_2$ gust components, and therefore would have been better represented by the Atassi function. However, since the gusts were composed of turbulent fluctuations, the corresponding $k_1$ reduced frequencies were high ($k_1 >> 1$). Thus, it can be concluded, based on the results of the experiments and simulations presented in this section, that even though these previous experiments generated flow conditions for the Atassi function, clear Sears trends were retrieved from the data because the $k_1$ values tested were sufficiently large.

%% file: 06_discussion_and_conclusions.tex
\section{Discussion and Conclusions}

It has been demonstrated that the fundamental difference between the Sears and Atassi functions lies in the Atassi parameter $k_2$. In practice, the validity of the functions depends on a correct representation of the character of the gust, specifically the Atassi parameters $\alphaG$ and $\epsilon$. These conclusions have been confirmed in experiments, where the gust conditions were generated using an active grid and characterized and controlled  to an unprecedented degree. The differences between the transfer functions are robust across the range of parameters taken into account by the models. Moreover, the difference between gusts with $k_2 = 0$ and $k_2 > 0$ is only apparent at low reduced frequencies, after which the data become insensitive to the type of normalization applied. This suggests that the horizontal gust modeled by the Atassi function significantly affects the gust response of the airfoil only at low reduced frequencies.

Additionally, these data show that the presence of small turbulent fluctuations on the airfoil leads to improved correspondence with the predictions of the analytical models. Experiments investigating the effect of gusts on momentum transfer through laminar and turbulent boundary layers would shed light on the physical mechanism responsible for this phenomenon.

The results of this study show that the correspondence with the Atassi function found by \cite{cordes_note_2017} stems not from the thickness and camber of the airfoil but rather from the character of the gusts themselves. The effects of thickness, camber, and mean angle of attack modeled by the Atassi function but not the Sears function, must therefore be decoupled from the gust character in order to be fully analyzed in an experimental context. At higher angles of attack and gust amplitudes, dynamic-stall flow phenomena become significant and cause the unsteady loads to deviate from the predictions of the Sears and Atassi functions. Studying these flow structures and their resulting load profiles would extend the findings of this study into the realm of real-world applications such as wind turbines and rotorcraft. Experiments along these lines were conducted in the scope of this project, and the results will be presented in a future article.

An obvious consequence of these results is that inflow conditions in active grid wind tunnels must be carefully scrutinized for possible (unintended) non-zero $k_2$ components, especially if gusts at low $k_1$ values are of interest. The present results reveal the impact of such components on airfoil gust response; however, their effect in other flow situations, e.g. on a vehicle, have yet to be studied in detail. A further implication is that additional effort in characterizing and tuning active grid generators may be necessary to avoid unwanted effects.

%% file: 07_acknowledgements.tex
\section{Acknowledgements}

The authors wish to acknowledge financial support of this study by the Deutsche Forschungsgemeinschaft through the Research Group PAK 780 ``Wind Turbine Load Control under Realistic Turbulent In-Flow Conditions". The group in Darmstadt would further like to acknowledge financial support by the Sino-German Center through the project TR 194/55-1 ``Flow Control for Unsteady Aerodynamics of Pitching/Plunging Airfoils". Nathan Wei was supported by the German-American Fulbright Commission with a grant in the Student Category during his stay at the TU Darmstadt.
Furthermore, the authors would like to greatly thank their respective workshop staff for the enduring and professional support throughout preparation and conducting of the measurement campaigns.